\documentclass[aps,prb,twocolumn,amsmath,amssymb,nofootinbib,superscriptaddress,floatfix]{revtex4-1}
\usepackage{graphicx,color}
\usepackage{color} 

\usepackage{bm} 
\usepackage{graphicx}
\usepackage{amssymb}
\usepackage{amsmath}
\usepackage{curves}
\usepackage{dcolumn} 

\usepackage[hypertex]{hyperref}

\begin{document}



\title{
Disorder-induced metal-insulator transitions
in three-dimensional topological insulators and superconductors
}


\author{Shinsei Ryu}
\affiliation{
Department of Physics, University of Illinois, 
1110 West Green St, Urbana IL 61801
            }
\affiliation{
Condensed Matter Theory Laboratory,
RIKEN, Saitama 351-0198, Japan
            }

\author{Kentaro Nomura}
\affiliation{
Correlated Electron Research Group (CERG), RIKEN-ASI, Wako 351-0198, Japan
            }

\date{\today}

\begin{abstract}
We discuss the effects of disorder in time-reversal invariant 
topological insulators and superconductors in three spatial dimensions.
For three-dimensional topological insulator in symplectic (AII) symmetry class,
the phase diagram in the presence of disorder and
a mass term, which drives a transition between trivial 
and topological insulator phases, 
is computed numerically by the transfer matrix method.
The numerics is supplemented by a field theory analysis
(the large-$N_f$ expansion where $N_f$ is the number of valleys or Dirac cones),
from which we obtain the correlation length exponent,
and
several anomalous dimensions at a non-trivial critical point
separating a metallic phase
and a Dirac semi-metal. 
A similar field theory approach is developed for
disorder-driven transitions 
in symmetry class AIII, CI, and DIII.
For these three symmetry classes, 
where topological superconductors 
are characterized by integer topological invariant, 
a complementary description is given 
in terms of the non-linear sigma model supplemented with
a topological term which is a three-dimensional 
analogue of the Pruisken term in the integer quantum Hall effect. 
\end{abstract}

\maketitle

\section{introduction}

A topological phase is a gapped quantum state of matter
with non-trivial quantum correlation (entanglement). 
Their key distinction from trivial states
a band insulator, say, lies in the fact that
they support stable gapless boundary modes
when terminated by a boundary. 
A systematic classification of topological insulators and superconductors
in all Altland-Zirnbauer symmetry classes
\cite{Zirnbauer96}
and in all dimensions was obtained, revealing  
mod 2 and 8 periodicities in dimensionality and symmetry classes.
\cite{Schnyder08,NewJ10,KitaevLandau100Proceedings}  
It was found in any dimensions
there exist topologically non-trivial insulators or superconductors
for five out of ten Altland and Zirnbauer symmetry classes.

In three spatial dimensions,
among the five symmetry classes is 
symplectic (spin-orbit) symmetry class (class AII), 
in which $\mathbb{Z}_2$ topological insulators can be realized.
\cite{Moore06, Roy3d, Fu06_3Da, Fu06_3Db,
hasanKaneReview10,
qiZhangReview10}
A number of materials 
with non-trivial $\mathbb{Z}_2$ topological number
have been discovered,
starting from Bismuth-Antimony alloys. 
(See, for a partial list of such materials,
Refs.\ \onlinecite{Hasan07,Xia2009,Chen09,Brune2011}.)

Topological superconductors that are possible in three dimensions
are all time-reversal symmetric; they are realized in
symmetry class CI (conserved SU(2) spin rotation symmetry), 
AIII (partially conserved SU(2) spin rotation symmetry),  
and DIII (no SU(2) spin rotation symmetry).
In these symmetry classes topologically 
different gapped states are characterized in the bulk by 
an integral, as opposed to $\mathbb{Z}_2$, 
topological number $\nu$ defined in terms of 
the spectral projector in momentum space. 
In these topological superconductors,
the Bogoliubov quasiparticles are fully gapped in the bulk
by the (mean field) pairing gap, and the system is a thermal insulator.
(When $z$-component of spin is a good quantum number as in 
symmetry class CI and AIII, 
spin transport can be discussed;
topological superconductors in these classes are 
a spin insulator in the bulk.)
On the other hand, when such 
three-dimensional (3d)
topological superconductors are 
terminated by a 2d surface, they can support gapless and 
stable Dirac (class AIII and CI) or Majorana (class DIII)
surface modes, the multiplicity 
(number of Dirac or Majorana cones) 
of which is specified by the bulk topological number $\nu$.
The fermionic quasiparticles in the B phase of ${ }^{3}$He 
(the BW state) realize an example of topological superconductor 
(superfluid) in class DIII.
Singlet BCS pairing models on the diamond and cubic lattices
that realize a topological superconductor in class CI
have also been proposed. 
\cite{diamond09,Hosur2010}


While topological insulators and superconductors
are stable against weak perturbations,  
with strong perturbations,
they can undergo a (continuous) quantum phase transition
into a different phase.
For example, if one can tune the chemical potential 
in the BW state, there is a transition between 
a trivial phase (strong pairing phase) and 
a topological phase (weak pairing phase).
Likewise in the quantum Hall effect (QHE),
strong enough disorder can also destroy the topological
insulators and superconductors, and drive a system,
through a continuous quantum phase transition,
into a metal or other topological/trivial phases.

This paper is devoted to study such 
quantum phase transitions among 
topological (electrical, thermal or spin) insulators,
trivial insulators, and a metal,
in symmetry class AII, 
and in three Bogoliubov-de Gennes (BdG) symmetry 
classes CI, AIII and DIII.
These quantum phase transitions can be induced by changing parameters 
that modify the band structure and wavefunctions, 
and also by introducing disorder;
both of these perturbations are assumed to be 
preserving discrete symmetries defining these symmetry classes.

\subsection{main results and outline of the paper}

\paragraph{symplectic (AII) symmetry class}

We start,
in Sec.\ \ref{effects of disorder in three-dimensional topological insulator in symmetry class AII},
by introducing a 3d Dirac-type Hamiltonian on the cubic lattice
in symplectic (AII) symmetry class,
which realizes a topological insulator,
a trivial insulator, and a phase transition between these two.
In Subsec.\ \ref{numerics},
the phase diagram in the presence of a mass term and disorder is
determined numerically by the transfer matrix method.
(For earlier studies, see, for example,
Refs.\ \onlinecite{Shindou2008,Shindou2010,TAI4,GoswamiChakravarty}.)
We observe that with disorder 
the phase boundary between topological and trivial phases
is renormalized, and in particular
the topological phase gets ``enlarged''.
This means,
as we pass through along a particular cut in the phase diagram, 
we can turn a trivial insulator into a topological
one by increasing disorder strength (``topological Anderson insulator''). 

The numerical analysis is supplemented 
with a field theory analysis 
(the large-$N_f$ expansion of a Gross-Neveu-type model 
where $N_f$ is the number of valleys or Dirac cones),
in Subsec.\ \ref{field theory AII},
from which we obtain the correlation length exponent,
several anomalous dimensions at non-trivial critical point,
and dynamical exponent.

\paragraph{BdG symmetry classes AIII, DIII and CI}

In time-reversal symmetric superconducting symmetry classes 
(class AIII, DIII and CI),
we will develop,
in Sec.\ \ref{effects of disorder in 
three-dimensional chiral topological superconductors},
a similar large-$N_f$ field theory approach
in terms of chiral Gross-Neveu- or Nambu-Jona-Lasinio-type models 
and their large-$N_f$ expansion.

While description in terms of these fermionic field theories 
are reliable for a region of the phase diagram 
where spin or thermal conductivity is relatively small
(in particular near the clean fixed point separating trivial and topological
superconductors), 
deep in a spin or thermal metallic phase, on the other hand,
a complementary description 
in terms of the non-linear sigma model (NL$\sigma$M) is possible
[Subsec.\ \ref{non-linear sigma models}].
For these three symmetry classes, where topological superconductors are
characterized by the integer topological invariant $\nu$, 
the NL$\sigma$Ms can be supplemented with a topological term 
which is of integer type (related to
$\pi_3(G/H)=\mathbb{Z}$ of the NL$\sigma$M target space $G/H$);
it is a higher dimensional analogue of the Pruisken term
in the QHE
\cite{Pruisken84, Levine84, Levine84b}.
We will compute the theta angle in the topological term 
in terms of microscopic parameters (the fermion mass
and the imaginary part of the fermion self-energy in
the self-consistent Born approximation).
While we do not have an analytical tool to analyze
the effects of the topological term on the nature of
Anderson metal-insulator transitions,
we will speculate on possible RG flows
in terms of diagonal spin or thermal conductivity
and the theta parameter.

\section{disorder in three-dimensional 
topological insulator in symmetry class AII}
\label{effects of disorder in three-dimensional 
topological insulator in symmetry class AII}

\subsection{Hamiltonian and symmetries}

Let us start by considering the following tight-binding Hamiltonian
\cite{Wilson74,Qi08}:
\begin{align}
H_0
&=
\sum_{{\bf x}}
\sum_{k=1,2,3}
\left[
\frac{{i}t}{2} c_{{\bf x}+{\bf e}_k}^{\dag}
\alpha^{\ }_{k}c^{\ }_{{\bf x}}
- \frac{r}{2}
c_{{\bf x}+{\bf e}_k}^{\dag}\beta c^{\ }_{{\bf x}} + \mathrm{h.c.}
\right]
\nonumber \\
&\quad
+ (m+3r)\sum_{{\bf x}}
c_{{\bf x}}^{\dag}\beta c_{{\bf x}}^{}, 
\label{lattice Dirac model AII}
\end{align}
where 
$c_{{\bf x}}^{\dag}, c_{{\bf x}}^{}$ represents the four-component 
fermion creation/annihilation operator defined on a site ${\bf x}$ on
the 3d cubic lattice spanned by
three orthogonal unit vectors ${\bf e}_{k=x,y,z}$, 
and
the $4\times 4$ gamma matrices in the Dirac representation are given by 
\begin{align}
\alpha_k =
\left(
\begin{array}{cc}
0 & \sigma_k \\
\sigma_k & 0
\end{array}
\right),
\,
\beta=\left(
\begin{array}{cc}
1 & 0 \\
0 & -1
\end{array}
\right),
\,
\gamma_5 =
\left(
\begin{array}{cc}
0 & 1\\
1 & 0
\end{array}
\right),
\label{gamma matrices}
\end{align}
with $\sigma_{k=1,2,3}$ being the standard Pauli matrices.
In momentum space,
the Hamiltonian is written as
$H_0
=
\sum_{\bf k} c^{\dag}_{\bf k}\mathcal{H}_0({\bf k})c_{\bf k}^{}
$, 
where $\mathcal{H}_0({\bf k})$ is given by
\begin{align}
&\quad
\mathcal{H}_0({\bf k})
=
\sum_{a=1}^3 
d_a({\bf k}) \alpha_a + d_4({\bf k})\beta, 
\nonumber \\
&\quad 
\mbox{with}
\quad 
\left\{
\begin{array}{ll}
\displaystyle
d_a({\bf k})
= t\sin k_a, \ \ \  (a=1,2,3), 
\\ 
\displaystyle
d_4({\bf k})
=
m+r
\sum_{a=1}^{3}
(1-\cos k_a),
\end{array}
\right. 
\end{align}
where ${\bf k}=(k_x,k_y,k_z)\in [-\pi, \pi)^3$. 
For arbitrary values of the parameters 
$t, m, r \in\mathbb{R}$,
the Hamiltonian is time-reversal symmetric,
$ ({i}\sigma_y) \mathcal{H}^*_0(-{\bf k})(-{i}\sigma_y)
=\mathcal{H}_0({\bf k})$. 
In the following we fix the Wilson parameter ($r$)
and the hopping amplitude ($t$) as
$r=1$ and $t=2$.

As we change $m$ in the Hamiltonian,
we can realize 
the trivial insulator,
the $\mathbb{Z}_2$ topological insulator,
and also a quantum critical point separating
trivial and topological insulators:
\begin{itemize}
\item
For $m < -6$, $-4< m<-2$ and $0< m$, 
$H_0$ realizes a trivial insulator whereas
\item
for $-6<m< -4$ and $-2<m< 0$, 
$H_0$ realizes a topological insulator. 
\end{itemize}
In the following we will mainly focus on
the region near $m=0$.

We add, to $H_0$, the following disorder potential
\begin{align}
V = 
\sum_{{\bf x}}
v_{{\bf x}}
c_{{\bf x} }^{\dag}
c_{{\bf x}}^{},
\end{align}
which is time-reversal symmetric,
and where the random variable $v_{{\bf x}}$ is distributed 
between $-W/2$ and $W/2$ according to the box-distribution.

When we are close to a quantum critical point,
say, at $m=0$, 
the tight binding Hamiltonian can be described 
in terms of its continuum limit, 
which takes the form of 3d massive Dirac Hamiltonian 
\begin{align}
\mathcal{H}
&=
\mathcal{H}_0+\mathcal{V},
\quad
\mathcal{H}_0
=
-{i}\partial_k \alpha_{k}
+ m \beta, 
\label{eq:dirac Hamiltonian AII}
\end{align}
where 
the summation over repeated indices
$k=1,2,3$
is implicit,
and 
$\mathcal{V}$ represents a disorder potential.
For any realization of disorder, 
the continuum Hamiltonian (\ref{eq:dirac Hamiltonian AII}) satisfies
time-reversal symmetry
\begin{align}
{i} \sigma_y \mathcal{H}^* (-{i}\sigma_y)
=
\mathcal{H}. 
\end{align}

The clean part of the Hamiltonian (\ref{eq:dirac Hamiltonian AII})
$\mathcal{H}_0$ realizes, depending on the sign of
the mass term $m\in \mathbb{R}$,
a topologically trivial and non-trivial insulator
in symmetry class AII, each of which is characterized by
the vanishing and non-vanishing, respectively,
of the $\mathbb{Z}_2$ invariant. 
When $m=0$ and in the absence of disorder, the spectrum is gapless whilst
the dc conductivity is zero, $\sigma_{xx}=0$. 
This should be compared with the ``universal'' finite 
conductivity in the 2d Dirac Hamiltonian. 
\cite{Ludwig94}

\subsection{numerics}
\label{numerics}

\paragraph{transfer matrix method}
In this section, we will study numerically
effects of disorder in the 3d Dirac model. 
Below, we use the transfer matrix
to study the random Hamiltonian $H=H_0+V$. 
To this end, we write the ``Dirac equation'' 
for the single-particle wavefunction
$\psi(\mathbf{x}, \sigma)=\psi(\mathbf{r}, \sigma)_n$
at a given energy $E$
in the following form: 
\begin{align}
\psi_{n+1}
&=
\frac{2}{1-t^2}(\beta+it\alpha_z)[\mathcal{H}^{(n)}-E]\psi_n
\nonumber \\
&\quad
           -\frac{1}{1-t^2}(\beta+i\alpha_z)(\beta-i\alpha_z)\psi_{n-1}.
\label{psi_n}
\end{align}
Here we decomposed the 3d spatial coordinates $\mathbf{x}$ in terms of
its $x$- and $y$- components $\mathbf{r}$
and 
its $z$-component $n$,
$\mathbf{x}=(\mathbf{r},n)$, and
${\cal H}^{(n)}$ is the Hamiltonian for the $n$th cross section.
As shown in Appendix \ref{transfer matrix},
this can be represented as 
\begin{align}
&
\begin{pmatrix}
\psi_{n+1} \\
\psi_{n}
\end{pmatrix}=M^{(n)}
\begin{pmatrix}
\psi_{n} \\
\psi_{n-1}
\end{pmatrix},\quad
  M^{(n)}=
\begin{pmatrix}
h^{(n)} & v \\
1 & 0
\end{pmatrix},
\nonumber \\
&\mbox{where}
\quad 
h^{(n)}
\equiv
\frac{-2}{t^2-1}(\beta+it\alpha_z)[\mathcal{H}^{(n)}-E].
\end{align}
The size of each slice with fixed $n$ is $L\times L$,
so that $M^{(n)}$ is a $4L\times 4L$ matrix. 
We will denote the length of the $z$-direction by $N$.

We study numerically
the dependence of the smallest Lyapunov exponent
of the transfer matrix $M$, 
as a function of the width $L^2$ of the
quasi-one dimensional system. 
Lyapunov exponents are 
self-averaging random variables
for an infinitely long quasi-one dimensional system, 
$N\to\infty$.

The eigenvalues of the Hermitian matrix
$M^{\dag} M$
are written as $\exp(\pm 2X_j)$ with
$0<X_1<X_2<\ldots<X_{4L}$.
The decay length $\xi^{\ }_L$ is then given by
\begin{equation}
\xi^{\ }_L=\lim_{N\to\infty}\frac{1}{NX_1}.
\end{equation}
The decay length $\xi^{\ }_{L}$
is a finite and self-averaging length scale
that controls the exponential decay of the Landauer conductance
for any fixed width $L$ of the infinitely long 
quasi-one dimensional system.
It is of course impossible to study infinitely long 
quasi-one dimensional lattice models
numerically and we shall approximate
$\xi^{\ }_{L}$
with $\xi^{\ }_{L,N}$ obtained from the Lyapunov exponents
of a finite but long quasi-one dimensional network model
made of $N$ slices. In our numerics we have set 
$N=1\times10^5\sim 5\times10^6$.

As shown by MacKinnon and Kramer,\cite{MacKinnon83}
criticality can be accessed from the
dependence of the normalized decay length
\begin{equation}
\Lambda_L :=\xi_L/L
\end{equation}
on the width $L$ of the quasi-one dimensional system. 
As usual, $\Lambda_L$ is required to be written as 
$\Lambda_L=f(L/\xi)$ on the basis of the single-parameter scaling hypothesis,
where $\xi$ is the (3d) localization length.
At the critical point the localization length diverges according to the power low
$\xi\sim |x-x_c|^{-1/\nu}$ upon tuning of a single microscopic parameter $x$ (disorder strength $W$ or mass $m$ in the following).

\begin{figure}[t]
\begin{center}
\includegraphics[width=0.4\textwidth]{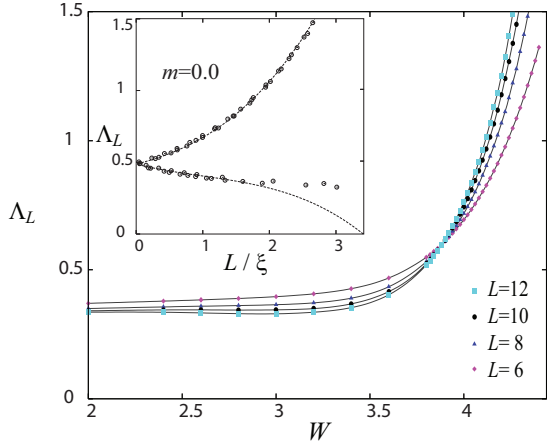}
\caption{
Normalized decay length $\Lambda_L=\Lambda/L$ as a function of disorder 
strength $W$
for $m=0.0$. Inset: A fit of the data shown in the main panel with the help of the one-parameter scaling ansatz in the vicinity of the critical point.
}
\label{xi-num}
\end{center}
\end{figure}

The numerically computed normalized decay length
is shown for $m=0$ and for different disorder strength in
Fig.\ \ref{xi-num}.
The 
plots
for different values of $L$ (=6, 8, 10, and 12)
cross at a single point $W\sim 3.9$,
signaling an insulator-to-metal transition. 
The inset shows $\Lambda_L$ as a function of $L/\xi$ with the help of the one-parameter scaling ansatz.
In a metallic region $\Lambda_L$ fits well with a typical metal-insulator transition behavior, while in the insulating phase $L$-dependence gets severer at weak disorder. 
This is so since in the insulating phase, the presence of the band gap would complicate the scaling analysis.
The fact that in the insulating side the data are not well fit by the scaling curve also makes the unambiguous determination of the critical exponent difficult. 
A similar problem was observed in the metal-insulator transition 
in the 2d quantum spin Hall systems.
\cite{QSHnum1,QSHnum2,QSHnum3}
A special effort on our 3d model will be made elsewhere.

\paragraph{structure of the phase diagram}

\begin{figure}[t]
\begin{center}
\includegraphics[width=0.4\textwidth]{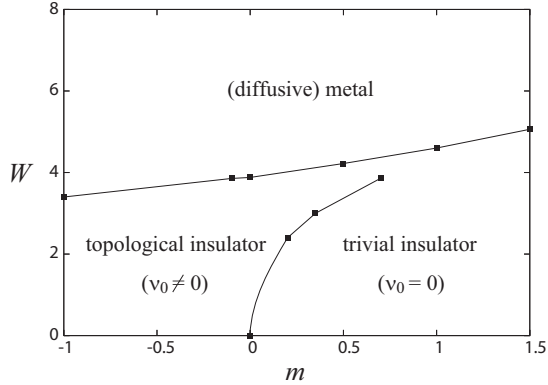}
\caption{
The numerically obtained phase diagram in the $(W, m)$ plane.
Lines are guide for eyes.
}
\label{phase-diag-num}
\end{center}
\end{figure}

In Fig.\ \ref{phase-diag-num}, 
the numerically determined phase diagram is shown.
Observe that as we increase disorder
the phase boundary separating 
trivial and topological insulators gets
shifted toward the large mass region, 
{\it i.e.}, the trivial insulator region
becomes topologically non-trivial as disorder strength increases.
This disorder-induced topological insulating phase has been found first 
in two dimensions\cite{TAI1,TAI2,TAI3,TAI3.1} 
and recently in three dimensions
\cite{TAI4,GoswamiChakravarty}, 
and thereby referred to as 
``topological Anderson insulator''. 
(However, one should note that the determination of 
the phase boundary between the two insulating phases is rather 
difficult numerically since in an insulator the $L$ dependence of the 
decay length is intrinsically small.) 
While it is not depicted here,
in the strong disorder limit, we expect there should be 
another phase boundary separating the metallic phase and
an insulator.

\subsection{field theory}
\label{field theory AII}

We now study the metal insulator transition
from field theory point of view. 
Our starting point
is the continuum limit of the lattice tight-binding model,
Eq.\ (\ref{eq:dirac Hamiltonian AII}),
\begin{align}
\label{Dirac model AII}
\mathcal{H}
&=
\mathcal{H}_0+\mathcal{V},
\quad
\mathcal{H}_0
=
-{i}\partial_k \alpha_{k}
+ m \beta. 
\end{align}
With time reversal symmetry,
the disorder potential 
$\mathcal{V}$ can be expanded in terms of 
$\sigma_{x,z}\tau_y$, 
$\sigma_{0}\tau_{0,x,z}$, 
and
$\sigma_y \tau_y$,
and hence parameterized by six real parameters characterizing 
disorder. 
Below, we assume, for simplicity, 
disorder potentials
are distributed according to 
a Gaussian distribution,
and
their variances are all equal, 
\begin{align}
P(V)
\propto 
\exp\left[
- 
\frac{1}{2g}
\mathrm{tr}_4\, \mathcal{V}\mathcal{V}^{\dag}
\right],
\end{align}
where $g$ is a parameter that characterizes the strength of disorder. 

\subsubsection{fermionic replica method}

We will use the fermionic replica method
for quenched disorder averaging. 
In the fermionic replica method, 
the discrete symmetry of the problem, time-reversal symmetry,
is implemented as a continuous symmetry: 
invariance under 
$\mathrm{O}(2N_r)\times \mathrm{O}(2N_r)$ 
rotations in the replica space
where $N_r$ is the number of replicas.  

In order to study Anderson localization physics,
we are after the properties of the single particle 
retarded (R) and advanced (A) Green's function,
$(\pm {i}0^+  -\mathcal{H})^{-1}$,
and in particular, the product thereof,
where $0^+ >0$ is a small imaginary part of the energy
needed to regularize poles.
For 
a product of the retarded (R) and advanced (A) Greens functions,
$\propto$ 
$({i}0^+-\mathcal{H})^{-1}
({i}0^++\mathcal{H})^{-1}
$, 
its generating function can be expressed as 
a fermionic functional integral,
$Z=
\int \mathcal{D}[\bar{\Psi}, \Psi] 
\exp( -\int d^3x \mathcal{L})$, 
where 
\begin{align}
\mathcal{L} = 
\bar{\Psi} \left(
{i}\mathcal{H} + 0^+ \Lambda 
\right)\Psi. 
\end{align}
Here, the total number of components 
of the fermion field $\bar{\Psi}, \Psi$ is 
\begin{align}
&\quad
\left(\mbox{Dirac}\right)
\otimes
\left(\mbox{TR}\right)
\otimes
\left(\mbox{R/A}\right)
\otimes
\left(\mbox{replica}\right)
\nonumber \\
&\quad
\quad
=
\sigma
\otimes
\tau
\otimes
\left(\mbox{R/A}\right)
\otimes
\left(\mbox{replica}\right)
=
16 N_{r} 
\end{align}
where
$\Lambda$ is a diagonal matrix which is $\pm 1$ for the
retarded/advanced sector, respectively.
The fermionic field $\Psi$ incorporates the time-reversal symmetry grading, 
and can be written as
\begin{align}
\bar{\Psi} 
=
\frac{1}{\sqrt{2}}
\left(
\bar{\chi}, -\chi^T {i} \sigma_y 
\right)_{\tau},
\, \, 
\Psi
=
\frac{1}{\sqrt{2}}
\left(
\begin{array}{c}
\chi \\
-{i}\sigma_y \bar{\chi}^T 
\end{array}
\right)_{\tau},
\label{tau grading}
\end{align}
where $\chi,\bar{\chi}$ are two-independent fermionic fields
(we have suppressed the replica indices.);
the subscript $(\cdots)_\tau$ means the block structure 
displayed in Eq.\ (\ref{tau grading}) represents 
the time-reversal symmetry grading.
It will prove convenient to rewrite Eq.\ (\ref{tau grading})
in the spin grading in which the block structure displays
subspaces with $\sigma=\pm$:
\begin{align}
\bar{\Psi} 
=
\frac{1}{\sqrt{2}}
\left(
\kappa^T \tau_{zy}, -\gamma^T \tau_{zy}
\right)_{\sigma},
\quad
\Psi
=
\frac{1}{\sqrt{2}}
\left(
\begin{array}{c}
-{i} \tau_{zy} \gamma \\
-{i} \tau_{zy} \kappa
\end{array}
\right)_{\sigma}, 
\end{align}
where 
\begin{align}
&-\kappa^T \tau_{yz}
:=
(\bar{\chi}_{\uparrow}, \chi_{\downarrow}^T),
\quad
\gamma^T \tau_{yz}
:=
(\bar{\chi}_{\downarrow}, -\chi_{\uparrow}^T),
\nonumber \\
&\tau_{zy}= (\tau_z-\tau_y)/\sqrt{2}. 
\end{align}
With these, the action can be written correspondingly as
\begin{align}
&
\quad 
\bar{\Psi}
\left(
{i} \mathcal{H} + \eta \Lambda 
\right)
\Psi 
\nonumber \\
&=
\frac{-{i}}{2}
\left(
-\kappa^T, \gamma^T
\right)_{\sigma}
\left(
{i} \mathcal{H} + \eta \Lambda 
\right)
\left(
\begin{array}{c}
\gamma \\
\kappa
\end{array}
\right)_{\sigma} 
\nonumber \\
&=
\frac{-{i}}{2}
\left(
\gamma^T, \kappa^T
\right)_{\sigma}
{i}\sigma_y 
\left(
{i} \mathcal{H} + \eta \Lambda 
\right)
\left(
\begin{array}{c}
\gamma \\
\kappa
\end{array}
\right)_{\sigma}. 
\end{align}
By introducing four component spinors by
\begin{align}
\psi =
\frac{1}{\sqrt{2}}
\left(
\begin{array}{c}
\gamma \\
\kappa
\end{array}
\right)_{\sigma},
\quad
\bar{\psi}
=
\psi^T C^{-1} 
=
\psi^T {i}\sigma_y , 
\end{align}
we end up with the functional integral
$Z =
\int \mathcal{D}[\bar{\psi},\psi] 
\exp\left(-
\int d^3x\, \mathcal{L}\right)
$
with the Lagrangian 
\begin{align}
\mathcal{L} &=
\sum_{a,b=1}^{4N_r}
\bar{\psi}_a 
\left(
{i} \mathcal{H} \delta_{ab} + 0^+ \Lambda_{ab}
\right)
\psi_b.
\end{align}
When $0^+=0$, 
the action is invariant under $\mathrm{O}(4N_r)$ rotations,
$
\kappa\to U \kappa
$,
$
\gamma\to U \gamma
$,
$U\in \mathrm{O}(4N_r)$.
In the metallic phase where a small imaginary part in the self-energy
is generated spontaneously, this symmetry is broken
down to $\mathrm{O}(2N_r)\times \mathrm{O}(2N_r)$.
The resulting Nambu-Goldstone mode is the diffuson 
in the metallic phase. 

We can now perform the disorder averaging according to 
$P(V)$, resulting in
the generating functional 
$Z =\int \mathcal{D}[\bar{\psi}, \psi ]
\exp\left(-\int d^3 x\, \mathcal{L}\right)$
with the Lagrangian
\begin{align}
\label{GN model AII}
\mathcal{L} &=
\sum_{a,b=1}^{4N_r}
\bar{\psi}_a
\left(
{i} \mathcal{H}_0 \delta_{ab} + 0^+ \Lambda_{ab}
\right)
\psi_b
-\frac{g}{2} 
\sum_{a,b}
\bar{\psi}_b
\psi_a \bar{\psi}_a
\psi_b.
\end{align}
When $0^+$ is zero,
this is a 3d version of 
the Gross-Neveu model
\cite{Gross_Neveu1974}
with $\mathrm{O}(4N_r)$ internal 
and $\mathrm{O}(N_f)$ flavor symmetries.

\subsubsection{saddle point} 

The disorder-induced four-fermion ``interaction''
can be decoupled by an auxiliary matrix field $W_{ab}$
leading to
$Z =
\int \mathcal{D}[\bar{\psi}, \psi] 
\int \mathcal{D} [W]
\exp\left(-\int d^3x\,\mathcal{L}\right)$, 
\begin{align}
\mathcal{L}
&=
\sum_{\iota=1}^{N_f}
\sum_{a,b=1}^{4N_r}
\bar{\psi}_{\iota a}
\left(
{i} \mathcal{H}_0 \delta_{ab} + 0^+ \Lambda_{ab}
+
 W_{ab} 
\right)
\psi_{\iota b}
\nonumber \\
&\quad
+
\frac{1}{2g} 
\mathrm{tr}_{4N_r} \, 
\left[
W W^T 
\right].
\label{eq: HS}
\end{align}
Here, we have introduced 
$N_f$ flavors of fermion fields
$\psi_{\iota a}$ ($\iota=1,\ldots, N_f$). 
Equation (\ref{eq: HS})
reduces to the situation we have been discussing when $N_f=1$,
whereas $N_f>1$ corresponds to the case with
$N_f$ ``valleys''. 
The replicated Lagrangian (\ref{eq: HS})
can be derived from a model similar to (\ref{Dirac model AII})
once we assume all the flavors 
are coupled equally by a disorder potential.
(See Appendix \ref{sec:
large-$N_f$ expansion for 
three-dimensional topological 
superconductors}.)
We will discuss the four-fermion ``interaction'' term 
in terms of the large $N_f$ expansion below. 

As a first step, we now look for a spatially homogeneous,
replica symmetric saddle point solution, by setting
\begin{align}
W_{ab} = \eta  \Lambda_{ab}.
\end{align}
This reduces the symmetry 
$\mathrm{O}(4N_r)\to
\mathrm{O}(2N_r)
\times
\mathrm{O}(2N_r)$. 
The self-consistency conditions for $\eta$ 
is given by
\begin{align}
&
\eta
\int^{\Lambda_{\mbox{\begin{tiny}UV\end{tiny}}}} \frac{d^3 k}{(2\pi)^3}
\frac{1}{k^2+m^2+\eta^2}
-
\frac{\eta}{4N_f g}
=
0,
\label{saddle}
\end{align}
where we have introduced
the ultra-violet cutoff $\Lambda_{\mbox{\begin{tiny}UV\end{tiny}}}$.
Anticipating spontaneous breaking 
of
$\mathrm{O}(4N_r)$ symmetry, 
we have set $0^+ =0$, 
as its sole role is to select the pattern of the residual
symmetry, $\mathrm{O}(2N_r)\times \mathrm{O}(2N_r)$.

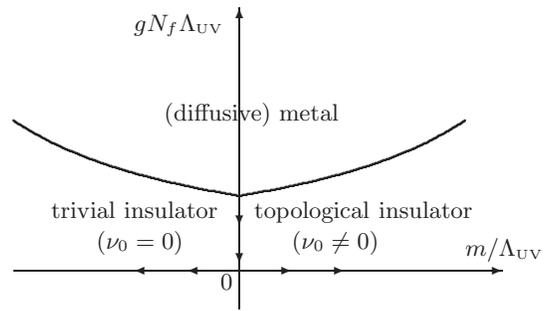
\begin{figure}[t] 
\begin{center}
\unitlength=10mm
\begin{picture}(6,4)(-3,-0.5)

\thinlines



\put(-3,0){\vector(1,0){6.5}}
\put(0,-0.5){\vector(0,1){4}}

\put(0,0){\line(0,1){1}}

\put(0,1.1){\vector(0,-1){0.5}}
\put(0,0.6){\vector(0,-1){0.5}}

\put(0,0)  {\vector(1,0){0.7}}
\put(0.7,0){\vector(1,0){0.7}}

\put(0,0)    {\vector(-1,0){0.7}}
\put(-0.7,0){\vector(-1,0){0.7}}

\thicklines

\put(-1.4,3.2){$g N_f \Lambda_{\mbox{\begin{tiny}UV\end{tiny}}}$}
\put(3,0.2){$m/\Lambda_{\mbox{\begin{tiny}UV\end{tiny}}}$}
\put(-0.25,-0.25){$0$}

\put(0.2,0.7){topological insulator}
\put(0.7,0.3){$(\nu_0 \neq 0)$}

\put(-2.5,0.7){trivial insulator}
\put(-1.9,0.3){$(\nu_0 = 0)$}

\put(-1,2){(diffusive) metal}

\curve(
	 0.000000, 1.000000,
	 0.500000, 1.090909,
	 1.000000, 1.200000,
	 1.500000, 1.333333,
	 2.000000, 1.500000,
	 2.500000, 1.714286,
	 3.000000, 2.000000)

\curve(
	 -3.000000, 2.000000,
	 -2.500000, 1.714286,
	 -2.000000, 1.500000,
	 -1.500000, 1.333333,
	 -1.000000, 1.200000,
	 -0.500000, 1.090909,
	 0.000000, 1.000000)

\end{picture}
\end{center}
\caption{
\label{fig: phase dirgram}
The mean field (self-consistent Born) phase diagram
near the Dirac point.
Arrows indicate the renormalization group flow. 
        }
\end{figure}

The resulting mean-field (or self-consistent Born) 
phase diagram is given in Fig.\ \ref{fig: phase dirgram}.
We identify the phase with $\eta \neq 0$ 
(the ordered phase with spontaneous $\mathrm{O}(4N_r)$ symmetry breaking)
as a (diffusive) metallic phase.
This metallic phase appears because disorder
creates states that fill the pseudo band gap of the Dirac fermions. 
On the other hand, there are two, trivial and topological,
insulating phases with $\eta=0$ which are separated by a phase boundary
which runs vertically at $m=0$.

In the lattice Dirac model
(\ref{lattice Dirac model AII}), 
a unitary transformation
$c^{\ }_{{\bf x}} \to {i}\gamma_5c^{\ }_{{\bf x}}$,
$c^{\dag}_{{\bf x}} \to -c^{\dag}_{{\bf x}}{i}\gamma_5$
relates 
two Hamiltonians, 
one with the parameter $(m,r)$
and the other with $(-m,-r)$; i.e.,
Hamiltonians with the mass $m$ and $-m$
are not unitary equivalent if $r$ is held fixed.
On the other hand,
the continuum Dirac model
(\ref{Dirac model AII})
and its replicated counterpart
(\ref{GN model AII})
are invariant under 
a $\mathbb{Z}_2$ transformation
$\psi_a\to \gamma_5\psi_a$,
$\bar{\psi}_a\to \bar{\psi}_a\gamma_5$,
$m\to -m$. 
For this reason, 
while the numerical phase diagram Fig.\ (\ref{phase-diag-num})
is asymmetric under $m\leftrightarrow -m$,  
the mean field phase diagram Fig.\ \ref{fig: phase dirgram}
is symmetric under $m\leftrightarrow -m$;
the ``topological Anderson insulator'' phase is not possible 
in the continuum model with linear dispersion.

\subsubsection{$1/N_f$-expansion}

Just at the Dirac point, 
the disorder, which appears as a four fermion interaction
in the generating functional,
is irrelevant, 
whereas the mass term is relevant,
from the power-counting.
The transition between trivial and topological insulators
is then described by the clean Dirac point. 
We will now turn our attention to the non-trivial fixed point
at $(m,g)= (0,g_c)$,
where $g_c\neq 0$ is the non-trivial solution to 
the saddle point equation 
Eq.\ (\ref{saddle}).
To include $1/N_f$ corrections systematically,
we consider the Lagrangian
\begin{align}
\mathcal{L}
&=
Z_{\psi}
\bar{\psi}_{\iota a}
(
\partial_{k} 
\gamma_k 
\delta_{ab} 
+
\Sigma \Lambda_{ab}
)
\psi_{\iota b}
\nonumber \\
&\quad
+
\frac{g}{\sqrt{N_f}} Z_{\psi}Z^{1/2}_W
\bar{\psi}_{\iota a}W_{ab} \psi_{\iota b}
+
\frac{1}{2} 
Z_{W}
W_{ab}W_{ab}, 
\label{eq: large N lagrangian AII}
\end{align}
where repeated flavor and replica indices are implicitly summed, 
and 
we have introduced Euclidean gamma matrices
$\gamma_k = \alpha_k$, \label{sec:
large-$N_f$ expansion for 
three-dimensional topological 
superconductors}
$g$ is the bare coupling 
and $Z_{\psi}, Z_{W}$ are wavefunction 
renormalization constants;
we have also rescaled the bosonic field $W_{ab}$ properly. 
These renormalization constants must be adjusted at each order in $1/N_f$ 
to keep vertex functions 
(in particular, the fermion propagator,
the fermion-boson vertex, and the boson propagator)
finite. 
(See, for example, Refs.\ \onlinecite{kocic,RWPreview,justinbook,moshe}.)

To leading order in $1/N_f$, as we have seen in 
the saddle point analysis, 
the expectation value of $W$ (=$\eta$)
is spontaneously generated if $g>g_c$:
the self-consistency equation is
\begin{align}
&
\eta
\int^{\Lambda_{\mbox{\begin{tiny}UV\end{tiny}}}} \frac{d^3 k}{(2\pi)^3}
\frac{1}{k^2+\eta^2}
-
\frac{\eta}{N_f D_{\gamma} g}
=
0,
\label{saddle 2}
\end{align}
where $D_{\gamma} (=4)$ is the dimensionality of
the Euclidean gamma matrices. 
The expectation value $\eta$
is a physical observable which sets a scale,
and hence independent of the UV cutoff. 
We will go to the non-trivial branch 
$g>g_c$ 
of the saddle point equation
(\ref{saddle 2})
for the most
part of our analysis below. 
This equation relates the bare coupling constant
$g$ to the UV cutoff and a physical scale 
$\eta$, and can be considered as a renormalization condition. 

The fermion propagator to leading order is given by
\begin{align}
\langle \psi_{\iota a}(k) \bar{\psi}_{\kappa b} (k) \rangle
&=
\delta_{\iota\kappa}
\mathcal{G}_{0,ab}(k),
\nonumber \\
\mathcal{G}_0^{-1}(k)
&=
2 Z_{\psi} ({i} k_i \gamma_i + \Sigma \Lambda_{ab}), 
\end{align}
where
we can choose 
$Z_{\psi}=1$ and $\Sigma=\eta$ to this order. 
The factor of 2 reflects the Majorana nature of the fermion fields.

\begin{figure}[t]
\centering
\includegraphics[width=0.45\textwidth]{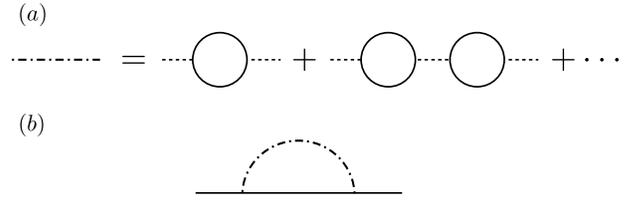}
\caption{
Some Feynman diagrams appearing in 
the $1/N_f$ expansion.
(a): summing fermion bubble diagrams
[Eq.\ (\ref{bosonic propagator1})].
(b): the $1/N_f$ expansion
correction to the fermion self-energy
[Eq.\ (\ref{ferminon self})].
\label{fig:diagrams}
}
\end{figure}

The boson two point function to leading order
can be computed
by summing over fermion bubble diagrams
(see Fig.\ \ref{fig:diagrams}), 
\begin{align}
\langle
W_{ab}(q)
W_{cd}(-q')
\rangle
=
\frac{8}{Z_{W} g^2 D_{\gamma}}
\mathcal{D}(q)
\delta_{q,q'}
\left(
\delta_{cb}\delta_{da}
+
\delta_{ca}\delta_{db}
\right),
\label{bosonic propagator1}
\end{align}
where $a,b,c,d=1,\ldots, 4N_r$,
$\delta_{q,q'}:=(2\pi)^3 \delta(q-q')$, 
and we have introduced 
\begin{align}
\mathcal{D}(q):= 
\frac{8\pi |q| }{
(q^2 + 4 \Sigma^2) \mathrm{atan}\,  (|q|/2\Sigma)
}. 
\label{bosonic propagator}
\end{align}
In the UV limit (i.e., at the critical point $g=g_c$), 
\begin{align}
\mathcal{D}(q)\to  
4/|q|,
\end{align}
and hence, 
by choosing the renormalization constant as
$Z_{W}g^2 = \Sigma^{-1}$, 
\begin{align}
&\quad
\langle
W_{ab}(q)
W_{cd}(-q')
\rangle
=
\frac{32}{ D_{\gamma} |q|}
\delta_{q,q'}
\left(
\delta_{cb}\delta_{da}
+
\delta_{ca}\delta_{db}
\right). 
\end{align}

The critical exponent $\nu$ is given, 
to leading order in $1/N_f$, by
$\nu = 1/(d-2)$ 
where $d=3$. 
We now compute 
next to leading order
(the order $1/N_f$) 
correction to the critical exponent $\nu$. 
First, 
we compute the wavefunction
renormalization and mass renormalization
to next-to-leading order.
The fermion propagator with 
$1/N_f$ corrections 
is given by
\begin{align}
\mathcal{G}^{-1}(k)
&=
2Z_{\psi} 
\Biggl\{
{i} k_i\gamma_i
\delta_{ab}
\left[
1+
\frac{16(4N_r +1)}{3\pi^2 N_f D_{\gamma} }
\ln
\frac{\Lambda_{\mbox{\begin{tiny}UV\end{tiny}}}}{|\eta|}
\right]
\nonumber \\
&\quad
+
\Sigma \Lambda_{ab}
\left[
1
- 
\frac{16}{\pi^2N_f D_{\gamma} }
\ln 
\frac{\Lambda_{\mbox{\begin{tiny}UV\end{tiny}}}}{|\eta|}
\right]
\Biggr\}.
\label{ferminon self}
\end{align}
From this, we determine
the counter terms as
\begin{align}
Z_{\psi}
&=
1- 
\frac{16(4N_r +1)}{3\pi^2 N_f D_{\gamma} }
\ln
\frac{\Lambda_{\mbox{\begin{tiny}UV\end{tiny}}}}{|\eta|}, 
\nonumber \\
\Sigma
&=
\eta
\left[
1
+
\alpha
\ln \frac{\Lambda_{\mbox{\begin{tiny}UV\end{tiny}}}}{|\eta|}
\right]
=
\eta 
\left(\frac{\eta}{\Lambda_{\mbox{\begin{tiny}UV\end{tiny}}}}\right)^{
-\alpha } 
\nonumber \\
\mbox{with}
&\quad
\alpha
= 
\frac{64(N_r +1)}{3\pi^2N_f D_{\gamma}}.
\end{align}

By definition, the critical exponent $\nu$
is determined from the divergence of the correction
length $\xi$ as a function of the distance from the critical point,
\begin{align}
t:=g^2 (g^{-2}_c-g^{-2}) \propto \xi^{- 1/\nu}. 
\end{align}
Note also that $\xi$ is inversely proportional to
$\eta$. 
On the other hand,
to this order, 
$t \propto \Sigma^{d-2}$. 
Combining these, 
\begin{align}
&
t \propto 
\eta^{
(d-2)
\left(
1-\alpha \right)}
=
\xi^{
-(d-2)
\left(
1-\alpha \right)}
\nonumber \\
\Rightarrow &
\quad 
\nu  =
\frac{1}{
(d-2)
\left(
1-\alpha \right)
}
=
\frac{(1+\alpha)}{
(d-2)}.
\end{align}
In particular, 
when $d=3$, $N_f=1$, $D_{\gamma}=4$, 
and in the replica limit $N_r\to 0$,
\begin{align}
\nu 
=
1+
\frac{ 16}{ 3\pi^2 }
\simeq 
1.54.
\end{align}
The result should be compared with
the critical exponent at 
the conventional Anderson transition 
in symplectic (AII) symmetry class,
which is estimated numerically as 
$\nu\sim 1.3-1.4$. 
(See Ref.\ \onlinecite{Asada2005} and references therein.)

\paragraph{dynamical exponent} 

So far we have been interested in static properties.
Some dynamical properties can also be computed within the large-$N_f$ expansion.
To this end, let us start from the imaginary-time action
\begin{align}
S &=
\int d\tau d^d x\, 
\sum^{N_f}_{\iota,\kappa=1}
\psi^{\dag}_{\iota} 
\left[
\left(
\partial_{\tau} + \mathcal{H}_0
\right)
\delta_{\iota\kappa}
+
\mathcal{V}_{\iota\kappa}
\right]
\psi^{\ }_{\kappa}.
\end{align}
Within the imaginary-time action,
electron-electron interaction terms such as 
$
(\lambda/2)
\int d\tau d^d x\, 
\big(
\sum^{N_f}_{\iota=1}
\bar{\psi}^{\ }_{\iota}\psi^{\ }_{\iota}
\big)^2
$
can be added. 
As before, 
disorder can be averaged by the fermionic replica method, 
and then the subsequent,
disorder-induced ``interaction'' term can be decoupled in terms of
the Hubbard-Stratonovich field $W(\tau,\tau',x)$ as 
\begin{align}
S &= 
\int d\tau d^d x\, 
\sum^{N_r}_{a=1}
\psi^{\dag}_a \left(
\partial_{\tau} + \mathcal{H}_0 
\right)
\psi_{a}
 \\
&\quad
+{i}
\int d^d x 
\int d\tau d\tau'
\sum_{a,b} 
\psi^{\dag}_a(\tau) 
W_{ab}(\tau,\tau')
\psi^{\ }_b(\tau')
\nonumber \\
&\quad
+
\frac{1}{2g}
\int d^dx\,
\int d\tau d\tau'
\sum_{ab}
W_{ab}(\tau,\tau',x)
W_{ba}(\tau',\tau,x). 
\nonumber 
\end{align}
This action can be studied by the saddle point method.
We assume the saddle point is spatially homogeneous and
replica symmetric, 
\begin{align}
W_{ab}(\tau,\tau',x)
=
X(\tau, \tau') \delta_{ab},
\end{align}
and $X(\tau, \tau')$ in frequency space is 
diagonal 
\begin{align}
X({i}\omega_{n},{i}\omega_{n'}) 
=
\delta(\omega_n-\omega_{n'})
X({i}\omega_n).
\end{align}

The electron Green function is given by,
at the disorder-dominated critical point, 
\begin{align}
\mathcal{G}^{-1}(\omega_n,k)
&=
-{i}\omega_n
+ \mathcal{H}_0(k)
+ {i}X({i}\omega_n)
\nonumber \\
&=
\mathcal{H}_0(k)
-{i}
C_d
|\omega_n|^{\frac{1}{d-1}}
\mathrm{sgn}\,\omega_n, 
\label{eq: dynamical}
\end{align}
where $C_d$ is a dimensionful numerical constant,
$ 
C_d=
\left\{
[2 |\sin (\pi d/2)|]/[\pi (d-2)]
\right\}^{1/(d-1)}
\Lambda^{(d-2)/(d-1)}_{\mbox{\begin{tiny}UV\end{tiny}}} 
$. 
From this, the dynamical exponent is 
\begin{align}
z=d-1. 
\end{align}
The Green's function (\ref{eq: dynamical})
should be contrasted with the non-relativistic case where
the disorder induced imaginary part of the self-energy 
within the self-consistent Born approximation
is given by 
\begin{align}
{i} X({i}\omega_n)
=
-
\frac{{i}}{2\tau}
\mathrm{sgn}\, \omega_n, 
\end{align} 
where 
$\tau= 1/[2 \pi \rho(\varepsilon_F)g]$ 
is the elastic scattering time
with
$\rho(\varepsilon_F)$
being the density of states at the Fermi energy
$\varepsilon_F$. 
\cite{BelitzKirkpatrick1994}

\section{disorder  in 
three-dimensional topological superconductors} 
\label{effects of disorder in 
three-dimensional chiral topological superconductors}

The Bogoliubov-de Gennes (BdG) Hamiltonian
describing the dynamics of fermionic quasiparticles 
in the superconducting (superfluid) phase realizes,
depending on its symmetry properties,
six out of ten Altland and Zirnbauer symmetry classes.
They can be summarized, 
in terms of how badly SU(2) spin rotation symmetry is broken, as follows:
A family (an ensemble) of  BdG Hamiltonians with completely broken 
spin rotation symmetry
is called symmetry class D (no TRS) or DIII (with TRS).
When one component ($z$-component, say) of spin is conserved, 
symmetry class A (no TRS) and AIII (with TRS) are realized.
\cite{Kunyang, Foster2008}
Finally, when full SU(2) symmetry is preserved, 
class C (no TRS) and CI (with TRS) are realized.

It is worth mentioning that historically symmetry class A and AIII 
have been discussed mainly in the context of electronic systems
(non BdG systems):
In the original paper by Altland and Zirnbauer,
class D, DIII, C, and CI are called ``BdG classes'', 
but not class A and AIII.
\cite{qcd}
Symmetry class A, or unitary class,
appears quite generally for electron systems which lack with 
TRS, and also with any other discrete symmetries.
Symmetry class AIII describes an electronic system
with broken TRS and with {\it sublattice} symmetry;
i.e., tight-binding Hamiltonians with bipartite hopping 
elements only.
While sublattice symmetry implies 
an invariance of the single-particle energy spectrum 
under reflection about zero energy, 
this is {\it not} related to 
PHS of BdG systems when a class AIII system
is interpreted as describing BdG quasiparticles. 
Below, when we speak about symmetry class AIII,
we will focus on its superconducting interpretation,
since sublattice symmetry in electronic systems 
requires fine-tuning,
while class AIII in BdG systems arises more naturally.

In three spatial dimensions,
there are three symmetry classes,
out of ten Altland and Zirnbauer symmetry classes, 
for which non-trivial topological phases
characterized by an integer topological invariant exist.
These are symmetry class CI, AIII and DIII. 
BdG Hamiltonians in these classes 
have two discrete symmetries,
TRS and PHS of different kind
and as a result they anticommute with a unitary matrix. 
For this reason, topological phases in these symmetry classes
are called chiral topological superconductors in Ref.\ \onlinecite{Hosur2010}. 

In this section, we will discuss effects of disorder in
these 3d topological superconductors. 
For the most part of this section,
we will mainly focus on symmetry class AIII.
The formalism for classes DIII and CI are largely parallel 
to class AIII, and will be discussed in Appendix.

\subsection{more descriptions of models}

\paragraph{BdG Hamiltonian}

The dynamics of quasiparticles deep inside a superconducting phase 
is described by a BdG Hamiltonian. 
On a lattice with $N$ sites, 
which is convenient for the purpose of discussing 
discrete symmetries of various kind, 
it can be written as follows:
\begin{equation}
H
=\frac{1}{2}
\big(
\begin{array}{cc}
\boldsymbol{c}^{\dag}, & \boldsymbol{c}
\end{array}
\big)
\mathcal{H}_4
\left(
\begin{array}{c}
\boldsymbol{c}^{\ } \\
\boldsymbol{c}^{\dag}
\end{array}
\right),\,\,
\mathcal{H}_4
=
\left(
\begin{array}{cc}
\Xi & \Delta \\
\Delta^{\dag} 
& -\Xi^{{T}}
\end{array}
\right),
\label{eq: BdG hamiltonian}
\end{equation}
where $\mathcal{H}_4$ is a $4N \times 4N$ matrix 
and $\boldsymbol{c}=
\left(\boldsymbol{c}_{\uparrow},\boldsymbol{c}_{\downarrow}\right)$.
[$\boldsymbol{c}$ and $\boldsymbol{c}^{\dag}$ 
can be either column or row vector
depending on the context.]
Because of
$\Xi=\Xi^{\dag}$ (hermiticity)
and
$\Delta=-\Delta^{{T}}$ (Fermi statistics),
the BdG Hamiltonian (\ref{eq: BdG hamiltonian})
satisfies PHS
\begin{align}
\mbox{[PHS]}:
\mathcal{H}^{\ }_4=-t_{x}\mathcal{H}^{{T}}_4 t_{x},
\quad
\label{eq: def PHS (triplet)}
\end{align}
where the Pauli matrix $t_{x,y,z}$ is acting on the particle-hole
grading. 
In addition to PHS, depending on physical systems,
TRS
\begin{align}
\mbox{[TRS]}
:
\mathcal{H}^{\ }_4= {i}s_y \mathcal{H}^{{T}}_4 (-{i}s_y),
\quad
\end{align}
and rotation symmetry for each spin rotation axis
\begin{align}
&\quad
\mbox{[SU(2) symmetry } (a=x,y,z)]:
\nonumber 
\\
&\quad
\qquad 
\big[\mathcal{H}_4,J_{a}\big]=0,
\quad
J_{a}:=
\left(
\begin{array}{cc}
s_{a} & 0 \\
0 & -s_{a}^{{T}}
\end{array}
\right),
\end{align}
can be satisfied,
where $s_{x,y,z}$ is the Pauli matrix acting on spin indices. 
Depending on their symmetry properties, 
BdG Hamiltonians (\ref{eq: BdG hamiltonian}) are classified into 
sub classes D, DIII, A, AIII, C, and CI.

\subsubsection{class DIII}

When both TRS and PHS
are preserved in the BdG Hamiltonian, 
the relevant symmetry class is
symmetry class DIII. 
Combining TRS and PHS, one can see that
a member of class DIII anticommutes with the unitary matrix $t_x \otimes s_y$,
\begin{align}
\mathcal{H}^{\ }_4 =
-t_x \otimes s_y \mathcal{H}_4 t_x\otimes s_y.
\end{align}
In this sense, class DIII Hamiltonians have
a chiral structure. It is sometimes convenient
to take a basis in which the chiral transformation,
which is $t_x \otimes s_y$ in the present basis,
is diagonal. In one of such bases,
a class DIII Hamiltonian takes on the form
\begin{align}
\mathcal{H}_4 =
\left(
\begin{array}{cc}
0 & D \\
D^{\dag} & 0
\end{array}
\right),
\quad
D = -D^T .
\label{eq: chiral basis for DIII}
\end{align}

A canonical example of the class DIII 
topological superconductor is 
the B phase of $^3$He,
\cite{Schnyder08,Roy08,Qi08,footnote1}
which is described, in momentum space,
by the following BdG Hamiltonian:
\begin{align}
H = 
\frac{1}{2}
\int d^3 k\,
\Psi^{\dag}({\bf k})
\mathcal{H}({\bf k}) 
\Psi({\bf k}), 
\end{align}
where 
$\Psi^{\dag}({\bf k})
= 
\big(
c^{\dag}_{\uparrow, {\bf k}},
c^{\dag}_{\downarrow, {\bf k}},
c^{\ }_{\uparrow,-{\bf k}},
c^{\ }_{\downarrow,- {\bf k}}
\big)
$
is the Nambu spinor 
composed of fermionic creation/annihilation 
operators ($c^{\dag}_{s, {\bf k}}/c^{\ }_{s, {\bf k}}$)
of a $^{3}$He atom 
with spin $s$ and momentum ${\bf k}$, 
and the kernel takes the following form:
\begin{align}
\mathcal{H}({\bf k})
=
\left(
\begin{array}{cc}
\xi({\bf k}) & \Delta({\bf k}) \\
\Delta^{\dag} ({\bf k})
& -\xi(-{\bf k})
\end{array}
\right).
\label{He3}
\end{align}
The matrix elements are given by 
\begin{align}
&\quad
\xi({\bf k})
=k^2/(2m) - \mu,
\nonumber \\
&\quad
\Delta({\bf k})
= 
\boldsymbol{d}({\bf k})\cdot \boldsymbol{s}
({i}s_y),
\quad 
\boldsymbol{d}({\bf k})
=
|\Delta| {\bf k},
\end{align}
where $m$ is the mass of a $^{3}$He atom, 
$\mu$ is the chemical potential,
and $|\Delta|$ is the amplitude of the pair potential. 
With the $\boldsymbol{d}$-vector pointing parallel 
to momentum, there is an isotropic gap everywhere on
the 3d fermi surface. 
The critical point at $\mu=0$ separates
topologically trivial ($\mu<0$)
and non-trivial ($\mu>0$)
phases, which are characterized by
the topological invariant $\nu=0$ and $\nu=1$, respectively. 
A recent surface transverse acoustic impedance measurement
reported in Ref.\ \onlinecite{murakawa09, Murakawa2010}
revealed the existence of the surface Majorana Dirac fermion 
mode on the surface of $^{3}\mathrm{He}$-B.

Many recently found non-centrosymmetric superconductors
also fall into class DIII.
\cite{noncentro10}
The theory of disordered topological superconductor in class DIII
which we will present is also relevant to these materials. 
A Majorana hopping model on the diamond lattice,
which realizes trivial ($\nu=0$) and non-trivial ($|\nu|=1$)
phases, as well as the Dirac Hamiltonian 
similar to Eq.\ (\ref{He3})
was discussed in 
Ref.\ \onlinecite{DiamondKitaev09}.

Below, we will consider effects of disorder around
this transition separating the weak-strong pairing phases, 
by perturbing the BdG Hamiltonian 
(\ref{He3})
by inhomogeneous, time-reversal invariant
potential $\mathcal{V}(x)$.
With a suitable unitary transformation, 
the full BdG Hamiltonian 
is written in terms of
the gamma matrices (\ref{gamma matrices})
as
\begin{align}
\mathcal{H}
=
|\Delta| (-{i} \partial_i) \alpha_i
+
\xi(-{i} \partial_i) \beta
+
\mathcal{V}(x).
\end{align}
We will henceforth set $|\Delta|=1$
and drop the $\mathcal{O}(k^2)$ term
in $\xi(\mathbf{k})$,
$\xi(\mathbf{k})\to -\mu$.

\subsubsection{class AIII}

Let us consider BdG Hamiltonians
which are invariant under rotations about the $z$- (or any fixed)
axis in spin space, 
$\left[ \mathcal{H}_4, J_{z} \right]= 0$. 
This implies that the Hamiltonian can be brought into the form:
\begin{align}
H
&=
\big(
\begin{array}{cc}
\boldsymbol{c}_{\uparrow}^{\dag}, &
\boldsymbol{c}^{\ }_{\downarrow}
\end{array}
\big)
\mathcal{H}_2
\left(
\begin{array}{c}
\boldsymbol{c}^{\ }_{\uparrow} \\
\boldsymbol{c}_{\downarrow}^{\dag} \\
\end{array}
\right),
\quad
\mathcal{H}_2
=
\left(
\begin{array}{cc}
\xi^{\ }_{\uparrow} & \delta \\
\delta^{\dag} & -\xi^{T}_{\downarrow}
\end{array}
\right),
\label{H_2}
\end{align}
where $\xi_{\sigma}$ and $\delta$ are an $N\times N$ matrix 
and 
$\xi^{\dag}_{\sigma}=\xi^{\ }_{\sigma}$.
Without further constraints,
this Hamiltonian is a member of class A
(unitary symmetry class).
If, in addition to conservation of the $z$-component of spin, 
we further impose TRS,
we obtain the conditions
$\xi^T_{\uparrow}
=
\xi^{\ }_{\downarrow}$ and 
$\delta =\delta^{\dag}$,
which can be summarized as
\begin{align}
r_y \mathcal{H}_2 r_y
=
-
\mathcal{H}_2,
\label{eq: class AIII condition in bdg H2}
\end{align}
where the Pauli matrices $r_{x,y,z}$ are acting on 
the particle-hole grading. 
It is also convenient to rewrite the Hamiltonian
by rotating the $r_{x,y,z}$ matrices by
$(r_x,r_y,r_z)
\Rightarrow
(r_x,-r_z,r_y).
$
In this basis,
the class AIII BdG Hamiltonian takes
on block off-diagonal form 
\begin{align}
\mathcal{H}_2
=
\left(
\begin{array}{cc}
0 & D \\
D^{\dag} & 0
\end{array}
\right). 
\end{align}

While we are not aware of any 
(lattice) BdG Hamiltonian realizing
topological superconductor in class AIII, 
an electronic 
lattice model of class AIII
with non-trivial topological charge $\nu=1$
is constructed in 
Ref.\ \onlinecite{Hosur2010}. 
In the continuum, the model can be described in terms 
of the following massive Dirac Hamiltonian: 
\begin{align}
\mathcal{H}_0
&= 
-{i}\partial_k \alpha_{k}
-{i}\beta \gamma^5 m_5.
\end{align}

\subsubsection{class CI}

If, in addition to $[\mathcal{H}_4, J_z]=0$, 
we further impose the full SU(2) rotation symmetry,
$\xi_{\sigma}$ and $\delta$ are constrained by
$\xi_{\downarrow}=\xi_{\uparrow}=:\xi$,
$\delta =\delta^T$.
This defines symmetry class C. 
Imposing both full SU(2) rotation and TR symmetries,
leads, 
in addition to 
$\xi_{\downarrow}=\xi_{\uparrow}$ and 
$\delta =\delta^T$, 
to
$\xi^* = \xi$,
and
$\delta^* =\delta$.
These conditions can be summarized as
\begin{align}
r_y \mathcal{H}^T_2 r_y
=
-\mathcal{H}^{\ }_2,
\quad
\mathcal{H}^*_2 =
\mathcal{H}^{\ }_2.
\end{align}
This defines symmetry class CI. 
Combining these two conditions
we can obtain a chiral symmetry,
$r_y \mathcal{H}_2 r_y
=
- \mathcal{H}_2
$.
By rotating the $r_{\mu}$ matrices as 
$(r_x,r_y,r_z)
\Rightarrow
(r_x,-r_z,r_y)
$, 
the class CI Hamiltonian takes
on block off-diagonal form 
\begin{align}
\mathcal{H}_2
=
\left(
\begin{array}{cc}
0 & D \\
D^{\dag} & 0
\end{array}
\right),
\quad
D=\delta-{i}\xi =D^T.
\label{eq: chiral rep for class CI}
\end{align}

In three dimensions, 
class CI is the only class which admits 
non-trivial topological state with SU(2) spin rotation 
symmetry. 
A BCS Hamiltonian on the diamond lattice
which belongs to
symmetry class CI
and realizes 
topological 
superconductor with $\nu=2$
was constructed in Ref.\ \onlinecite{diamond09}.
A model on the cubic lattice was also constructed\cite{Hosur2010},
and it was shown that a defect in such 
topological SC accumulate a spin 1/2 degree of
freedom at its core.

\subsection{continuum Dirac model and fermionic replica method}

We now start our discussion on the effects of
disorder on topological superconductors. 
Let us start by considering, 
as a model of a 3d topological superconductor in class AIII, 
a 3d massive Dirac Hamiltonian 
\begin{align}
\mathcal{H}
&= 
\mathcal{H}_0+\mathcal{V},
\nonumber \\
\mathcal{H}_0
&= 
-{i}\partial_k \alpha_{k}
-{i}\beta \gamma^5 m_5,
\nonumber \\
\mathcal{V}
&= 
\left(
\begin{array}{cc}
0 & D_I(x)\\
D^{\dag}_I(x) & 0
\end{array}
\right),
\label{eq:dirac Hamiltonian}
\end{align}
where
$D_I(x)$ in $\mathcal{V}$ represents a disorder potential.
For any value of the mass $m_5$
and for any realization of disorder, 
the Hamiltonian (\ref{eq:dirac Hamiltonian}) satisfies
chiral symmetry
\begin{align}
\beta  \mathcal{H} \beta
=
-\mathcal{H}.
\label{eq: chiral symmetry}
\end{align}
As discussed in Ref.\ \onlinecite{Schnyder08},
$\mathcal{H}_0$ realizes, depending on the sign of
the mass term $m_5\in \mathbb{R}$,
a topologically trivial and non-trivial superconductor 
in class AIII, each of which is characterized by
the vanishing and non-vanishing, respectively,
of the winding number $\nu$ defined for the projector
introduced in Ref.\ \onlinecite{Schnyder08}.
We assume $D_I(r)$ is maximally random
according to the distribution
$
\propto 
\exp
[
-\mathrm{tr}_2 (D^{\ }_I D_I^{\dag})/(2g)
].
$

To discuss effects of disorder,
it is convenient to introduce
the generating function,
$Z = 
\int\mathcal{D}[\psi^{\dag}_a,\psi^{\ }_a]
\exp\big(-\int d^3 x\,\mathcal{L}\big)$,
for the retarded Green's functions,
where
$\psi^{\dag}_a,\psi^{\ }_a$ with $a=1,\cdots, N_{r}$
is a replicated fermionic variable,
and 
\begin{align}
\mathcal{L} 
&= 
\sum^{N_{r}}
_{a=1}
\psi^{\dag}_a
\left(
{i}\mathcal{H}
+0^+
\right)
\psi_a.
\end{align}
We have introduced $N_{r}$ replicas for the quenched disorder averaging,
with $N_{r}\to 0$ at the end of calculation.
The $0^+>0$ is introduced for the convergence of the functional integral.
The Lagrangian enjoys an $\mathrm{U}(N_{r})$ symmetry 
\begin{align}
\psi^{\dag} \to \psi^{\dag}
T^{\dag},
\quad
\psi \to T \psi,
\quad T\in \mathrm{U}(N_{r}).
\end{align}
Due to the chiral symmetry of class AIII, there is an additional
(axial) $\mathrm{U}(N_{r})$ symmetry when $0^+=0$,
\begin{align}
\psi^{\dag} \to \psi^{\dag}
e^{+{i}\beta \theta},
\quad
\psi \to e^{+{i}\beta \theta} \psi,
\quad
\theta \in \mathrm{u}(N_{r}).
\end{align}
Averaging over disorder gives 
\begin{align}
\mathcal{L} &= 
\sum_a
\bar{\psi}_a
\left(
\gamma^k \partial_k
-{i}
0^+ \gamma^0
-{i}
m_5 \gamma^5
\right)
\psi_a
\nonumber \\
&\quad
-\frac{g}{2}
\sum_{a,b}
\left(
\bar{\psi}_a \psi_b
\bar{\psi}_b \psi_a
-
\bar{\psi}_a \gamma^0 \psi_b
\bar{\psi}_b \gamma^0 \psi_a
\right), 
\end{align}
where $\bar{\psi}=\psi^{\dag}\beta {i}$
and $\gamma^k=- {i}\beta \alpha^k $.
[Observe that we introduced 
$\gamma^k$ differently from
the case of symmetry class AII, 
Eq.\ (\ref{eq: large N lagrangian AII}). 
For symmetry class AIII, DIII, and CI,
we find this convention is more convenient.]
This is a 3d version of 
the chiral Gross-Neveu or
the Nambu-Jona-Lacinio model
\cite{Nambu61}
perturbed by a mass term.

The \textcolor{blue}{quartic} interactions among fermions
can be decoupled by 
the Hubbard-Stratonovich transformation
with two Hermitian matrices $W$ and $V$,
\begin{align}
\mathcal{L}
&= 
\sum_{a}
\bar{\psi}_a
\left(
\gamma^k \partial_k 
-{i}
0^+ \gamma^0 
-{i}
m_5 \gamma^5 
\right)
\psi_a
\nonumber \\
&\quad
+
\sum_{a,b}
\bar{\psi}_a
\left(
W_{ab}
+
{i}\gamma^0V_{ab}
\right)
\psi_b
\nonumber \\
&\quad
+\frac{1}{2g}
\mathrm{tr}_{N_r}\left(
W^2+V^2
\right).
\label{AIII GN anction with HS field}
\end{align}

There are similar Dirac type Hamiltonians 
describing 3d topological insulators in class CI and DIII,
and the following mean-field and large-$N_f$ analysis are completely
parallel for these classes. 
In the large-$N_f$ calculations,
the nature of Nambu-Goldstone fluctuations 
are different among symmetry classes AIII, DIII, and CI: 
Roughly, 
one needs to replace 
$\mathrm{U}\to \mathrm{O}$ for class DIII,
and 
$\mathrm{U}\to \mathrm{Sp}$ for class CI.

\subsection{large $N_f$-analysis}

We now look for a spatially homogeneous,
replica symmetric saddle point solution,
by setting
\begin{align}
W_{ab} = w \delta_{ab},
\quad
V_{ab} = v \delta_{ab}.
\label{saddle AIII}
\end{align}
It is enough to look for a saddle point with
$v\neq 0$ and $w=0$, 
since there is a symmetry that relates 
a saddle point
with $w=0$ and $v\neq 0$
and 
a saddle point
with $w\neq 0$ and $v = 0$. 
The resulting mean-field (self-consistent Born) 
phase diagram is essentially the same as
Fig.\ \ref{fig: phase dirgram}. 
(We need to replace $m\to m_5$). 
We identify the phase with $v\neq 0$ 
(the ordered phase of the NL$\sigma$M)
as a (diffusive) metallic phase.
This metallic phase appears because the disorder
creates states that fill the band gap.
While it is not depicted here,
in the large $g$ limit, there should be
an Anderson insulating phase in class CI and class DIII.
On the other hand, in class AIII,
it is not necessary to have an Anderson insulator
in the large $g$ limit 
since the beta function in $(2+\epsilon)$ dimensions is positive
to all order in perturbation theory, 
\cite{Gade91-93} 
as known in the context of 
the random hopping (flux) model in one and two dimensions 
where there is no Anderson localization. 
For a numerical study of the 3d random hopping model in class AIII,
See Ref.\ \onlinecite{Garcia-Garcia2006}. 
(This result should be considered, however, as the case with 
zero topological angle, $\theta=0$, while our system is located at 
$\theta = \pi/2$ when $m_5=0$; see below for more discussion.)

Just at the Dirac point, 
the disorder, which appears as a four fermion interaction
in the generating function,
is irrelevant, 
whereas the mass term is relevant,
from the power-counting.
The transition between trivial and topological insulators
is then described by the clean Dirac point.

In the large-$N_f$ expansion,
the model can be renormalized in a similar way 
as the class AII Gross-Neveu model
(\ref{eq: large N lagrangian AII}).
We introduce 
wavefunction renormalizations
$Z_{\psi}$ and $Z_{W}$ and consider 
the Lagrangian
\begin{align}
\mathcal{L}
&=
Z_{\psi}
\bar{\psi}_{\iota a}
(
\partial_{k} 
\gamma_k 
\delta_{ab} 
+
\Sigma 
)
\psi_{\iota b}
\nonumber \\
&\quad
+
\frac{g}{\sqrt{N_f}} Z_{\psi}Z^{1/2}_W
\bar{\psi}_{\iota a}
\left(
W_{ab} 
+
{i} \gamma^5 
V_{ab} \right)
\psi_{\iota b} 
\nonumber \\
&\quad
+
\frac{1}{2} 
Z_{W}
\left(
W_{ab}W_{ab}
+
V_{ab}V_{ab}
\right), 
\label{eq: large N lagrangian AIII}
\end{align}
where we have rescaled bosonic fields properly. 
To leading order, the fermion propagator
and the boson propagator
are given by 
\begin{align}
\langle \psi_{\iota a}(k) \bar{\psi}_{\kappa b} (k) \rangle
&=
\delta_{\iota\kappa}
\delta_{ab}
Z^{-1}_{\psi}
({i} k_i \gamma_i + \Sigma), 
\nonumber \\
\langle
W_{ab}(q) W_{cd}(-q')
\rangle
& =
\langle
V_{ab}(q) V_{cd}(-q')
\rangle
\nonumber \\
&
=
\frac{4}{Z_W g^2 D_{\gamma} }
\mathcal{D}(q^2)
\delta_{q,q'} 
\delta_{cb}
\delta_{da}.
\end{align}

As in the class AII Gross-Neveu model, 
several critical exponents 
can be readily computed within the large-$N_f$ approach.
For example, the anomalous dimensions 
of a fermion $\psi$ to leading order in $1/N_f$
is given by
\begin{align}
\eta_{\psi}
=
\alpha
=
\frac{16 N_r }{ 3\pi^2 N_f D_{\gamma} }
\quad
\mbox{for class AIII},
\end{align}
where the fermion Green's function behaves as
$\mathcal{G}(x)
\sim
|x|^{-2-\eta_\psi}$. 
In the replica limit, this leading order correction is vanishing. 
Similarly, 
the anomalous dimension of
the fermion mass is given by
\begin{align}
\eta_{\bar{\psi}\psi}
=
\frac{32 N_r }{ 3\pi^2N_f D_{\gamma}}, 
\label{mass dim AIII}
\end{align}
and the anomalous dimension of the ``$\gamma_5$'' fermion mass is given by
\begin{align}
\eta_{\bar{\psi}\gamma_5\psi}
=
\frac{32N_r}{ \pi^2 N_f D_{\gamma}} . 
\label{gamma5 mass dim AIII}
\end{align}

The similar calculations go through for class DIII, and CI.
For symmetry class DIII,
\begin{align}
\eta_{\psi}
=
\frac{64 N_r }{ 3\pi^2N_f D_{\gamma}}, 
\label{mass dim DIII}
\end{align}
and for the anomalous dimension of
the fermion mass is given by
\begin{align}
\eta_{\bar{\psi}\psi}
=
\frac{256 \left(N_r-3\right) }{3 \pi^2 N_f D_{\gamma}}. 
\end{align}

\subsection{non-linear sigma models}
\label{non-linear sigma models}

\begin{ruledtabular}
\begin{table}[t!]
\begin{center}
\begin{tabular}{ccc}
symmetry & NL$\sigma$M & 3d topological \\
class& target space & term
\\ \hline
A    &  
$\mathrm{U}(2N)/\mathrm{U}(N)\times \mathrm{U}(N)$  & 
\\ 
AI   &  $\mathrm{Sp}(2N)/\mathrm{Sp}(N)\times \mathrm{Sp}(N)$ &
\\ 
AII    & $\mathrm{O}(2N)/\mathrm{O}(N)\times \mathrm{O}(N)$  &

\\ \hline
AIII    & $\mathrm{U}(N)\times \mathrm{U}(N)/ \mathrm{U}(N)$  &
 $\mathbb{Z}$ 
\\ 
BDI   & $\mathrm{U}(2N)/ \mathrm{Sp}(N)$  & 
\\ 
CII    & $\mathrm{U}(2N)/ \mathrm{O}(2N)$  
& $\mathbb{Z}_2$ 
\\ \hline
D    & $\mathrm{O}(2N)/ \mathrm{U}(N)$  &
\\ 
C    & $\mathrm{Sp}(N)/ \mathrm{U}(N)$  & $\mathbb{Z}_2$
\\ 
DIII &   $\mathrm{O}(N)\times \mathrm{O}(N)/\mathrm{O}(N)$  & $\mathbb{Z}$
\\ 
CI   & $\mathrm{Sp}(N)\times \mathrm{Sp}(N)/\mathrm{Sp}(N)$  & $\mathbb{Z}$
\\ 
\end{tabular}
\caption{
\label{tab: nlsm}
Fermionic replica NL$\sigma$M target spaces and
possible topological terms in
three dimensions.
For a similar table for two spatial dimensions, 
see Ref.\ \onlinecite{Fendley00-01} .
}
\end{center}
\end{table}
\end{ruledtabular}

Deep in the diffusive metallic phase,
physics at large distances should be described by
the NL$\sigma$M
defined on
$\mathrm{U}(N_{r})$
(AIII),
$\mathrm{O}(2N_{r})$
(DIII),
and
$\mathrm{Sp}(N_{r})$
(CI),
respectively.
The coupling constant of the NL$\sigma$M
plays the role of the longitudinal conductivity $\sigma_{xx}$. 
(To be more precise, in superconductors,
$\sigma_{xx}$ corresponds to either 
spin or thermal conductivity.)
In addition, one should note here that
the NL$\sigma$M for these classes AIII, DIII, and CI can 
support
a topological term in three dimensions:
For the NL$\sigma$M target manifolds $G$,
where $G=\mathrm{U}(N_r)$ (AIII),
$G=\mathrm{O}(2N_r)$ (DIII), and
$G=\mathrm{Sp}(N_r)$ (CI),
the homotopy group is given by
$\pi_3(G)
=
\mathbb{Z}$.

The NL$\sigma$M and the topological term
can be obtained in the following way
(we will focus on symmetry class AIII as an example,
but a similar derivation should be possible
for class DIII and CI):
a saddle point solution in Eq.\ (\ref{saddle AIII})
breaks the global 
$\mathrm{U}(N_{r})\times\mathrm{U}(N_{r})$
symmetry down to
$\mathrm{U}(N_{r})$.
One then expects the long wave length physics
are described in terms of the Nambu-Goldstone modes associated with
the symmetry breaking.
With the inclusion of the Nambu-Goldstone modes
(=fluctuations around the saddle point), the action is 
\begin{align}
\mathcal{L} =
\bar{\psi}_a
\left[
\big(
\gamma^k \partial_k 
+m_5  \gamma^5 
\big)\delta_{ab}
+
q\gamma^0
\big(
e^{2{i}\gamma^0 \theta(x)}
\big)_{ab}
\right]
\psi_b,
\label{fermion + NG boson}
\end{align}
where the slow variations represented by
$\exp(2{i}\gamma^0 \theta(x))$
are the Nambu-Goldstone modes in the NL$\sigma$M,
\begin{align}
e^{2{i}\gamma^0 \theta(x)}
=
\left(
\begin{array}{cc}
U & 0 \\
0 & U^{\dag}
\end{array}
\right),
\quad
U \in \mathrm{U}(N_{r}).
\end{align}
The Lagrangian (\ref{fermion + NG boson})
is obtained from Eq.\ (\ref{AIII GN anction with HS field})
by freezing the longitudinal fluctuations but keeping
the transverse fluctuations (i.e., the Nambu-Goldstone modes).
The longitudinal fluctuations are gapped and can safely be integrated
over, which can renormalize parameters in the effective action. 
(In symmetry class AIII, it can also generate the so-called Gade term.
\cite{Gade91-93}
We do not discuss, however, the effect of the Gade term in this section.)
The effective action $S_{\mathrm{eff}}[U]$
for the Nambu-Goldstone modes
is then obtained by integrating over fermions,
\begin{align}
Z= 
\int \mathcal{D}\left[U\right]
\mathcal{D}\left[\bar{\psi},\psi\right]
e^{-S}
=
\int \mathcal{D}\left[U\right]
e^{-S_{\mathrm{eff}}[U]
}.
\end{align}
Expanding the effective action in powers
of $U$ and its gradient,
and with the Pauli-Villars regularization, 
\cite{Abanov2000}
the effective action (the NL$\sigma$M action) is given by
\begin{align}
S_{\mathrm{eff}}[U]
&=
\frac{\pi \sigma_{xx}}{4}
\int d^3 x\,
\mathrm{tr} \left(
\partial_j U^{\dag} \partial_j U
\right)
+
{i}
\theta
\Gamma [U], 
\end{align}
where the topological term 
\begin{align}
\Gamma [U]
=
\int \frac{d^3x}{24\pi^2} \epsilon^{ijk} 
\mathrm{tr} \left(
U^{\dag} \partial_i U
U^{\dag} \partial_j U
U^{\dag} \partial_k U
\right)
\end{align}
is an integer for any field configuration.
The $\theta$ angle (topological angle) 
is given
by (see Appendix 
\ref{non-linear sigma models and topological angle})
\begin{align}
\frac{\theta}{\pi}
&=
-\frac{m^3_5 }{ |M|^3}
+\frac{m^3_R}{|m_R|^3}
-\frac{9 m^{\ }_5 m^2_0 }{4 |M|^3}, 
\nonumber \\
M^2&:=m^2_5+m^2_0,
\end{align}
where $m_0$ is the imaginary part of the self-energy
(within the self-consistent Born approximation),
and $m_R$ is a mass introduced by the Pauli-Villars
regularization.  
Notice that 
the topological angle is $\pi$ when $m_5$ is $0$ (Dirac point)
for arbitrary $m_0$.

\subsubsection{possible two-parameter scaling phase diagram}

The presence of the topological term is a direct manifestation 
of the fact that the metallic phase is adjacent to a topological 
superconducting phase. 
The similar topological term in two dimensions, 
the Pruisken term in the QHE,  
has a significant implication
on the nature of Anderson localization physics,
and, in fact, is crucial for the existence of the QHE,
and for the plateau transition.  
The role of such Pruisken term
in topological superconductors with broken time-reversal symmetry 
in two dimensions has been discussed
in symmetry class A, D, and C. 
\cite{SenthlFisherBalentsNayak1998,
SenthilFisher2000,
SenthilMarstonFisher}

It is not clear what the role is played by the topological term
in 3d NL$\sigma$Ms. 
While controlled calculations 
near metal-insulator transitions
and in the presence of topological terms 
are in general difficult,
below we combine 
the RG flow of the NL$\sigma$M for large $\sigma_{xx}$
with the RG flow around the clean Dirac point from the $1/N_f$-expansion, 
to speculate the general structure of the phase diagram. 
\cite{Ghaemi2010}
As in the QHE, 
the phase diagram and the RG flow 
may be phrased in terms of 
the two parameters of the 3d NL$\sigma$M;
the NL$\sigma$M coupling constant $1/\sigma_{xx}$ 
and the topological angle $\theta$.
\cite{comment}

\begin{figure}[t]
\centering
\includegraphics[width=0.23\textwidth]{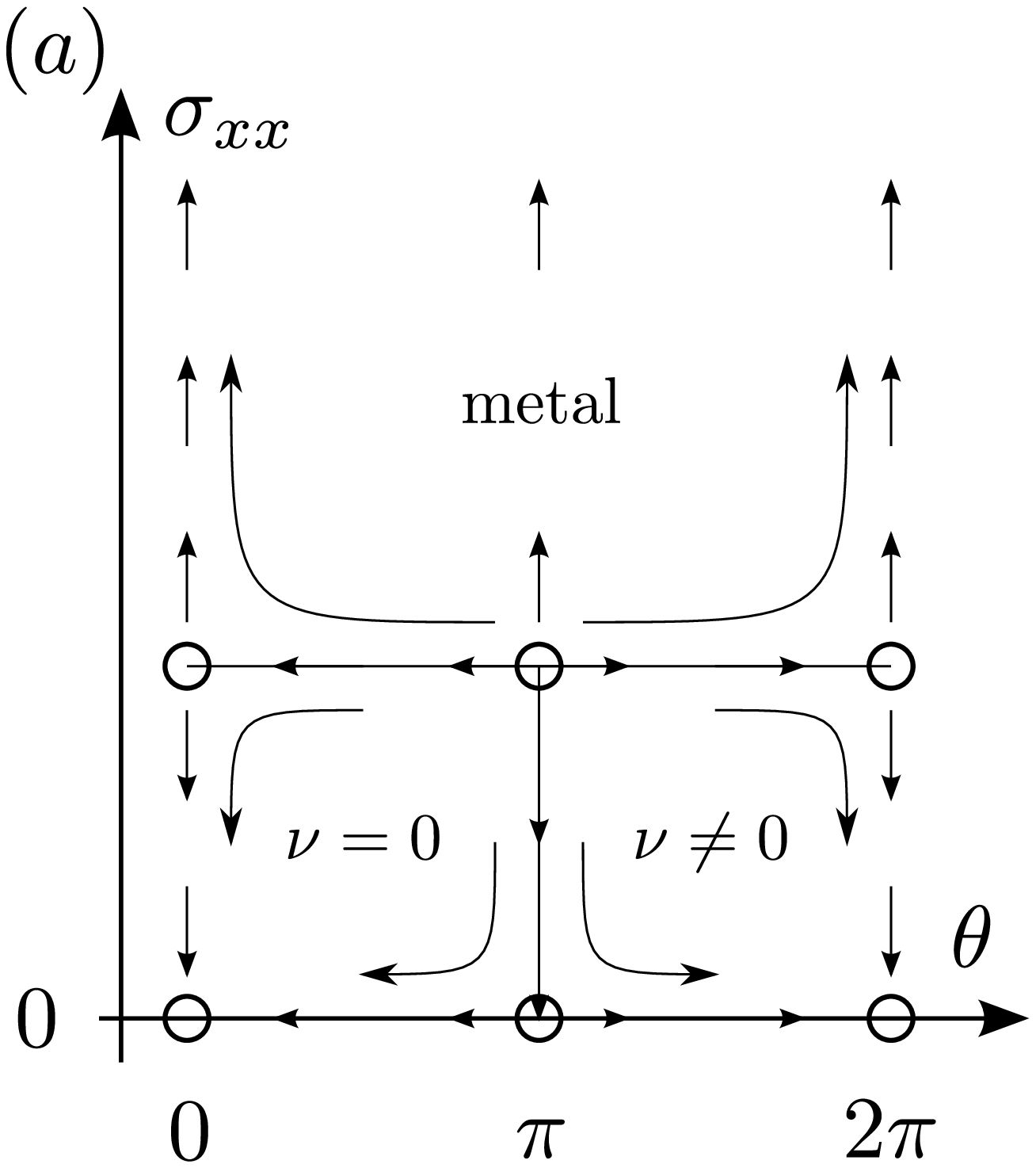}
\includegraphics[width=0.23\textwidth]{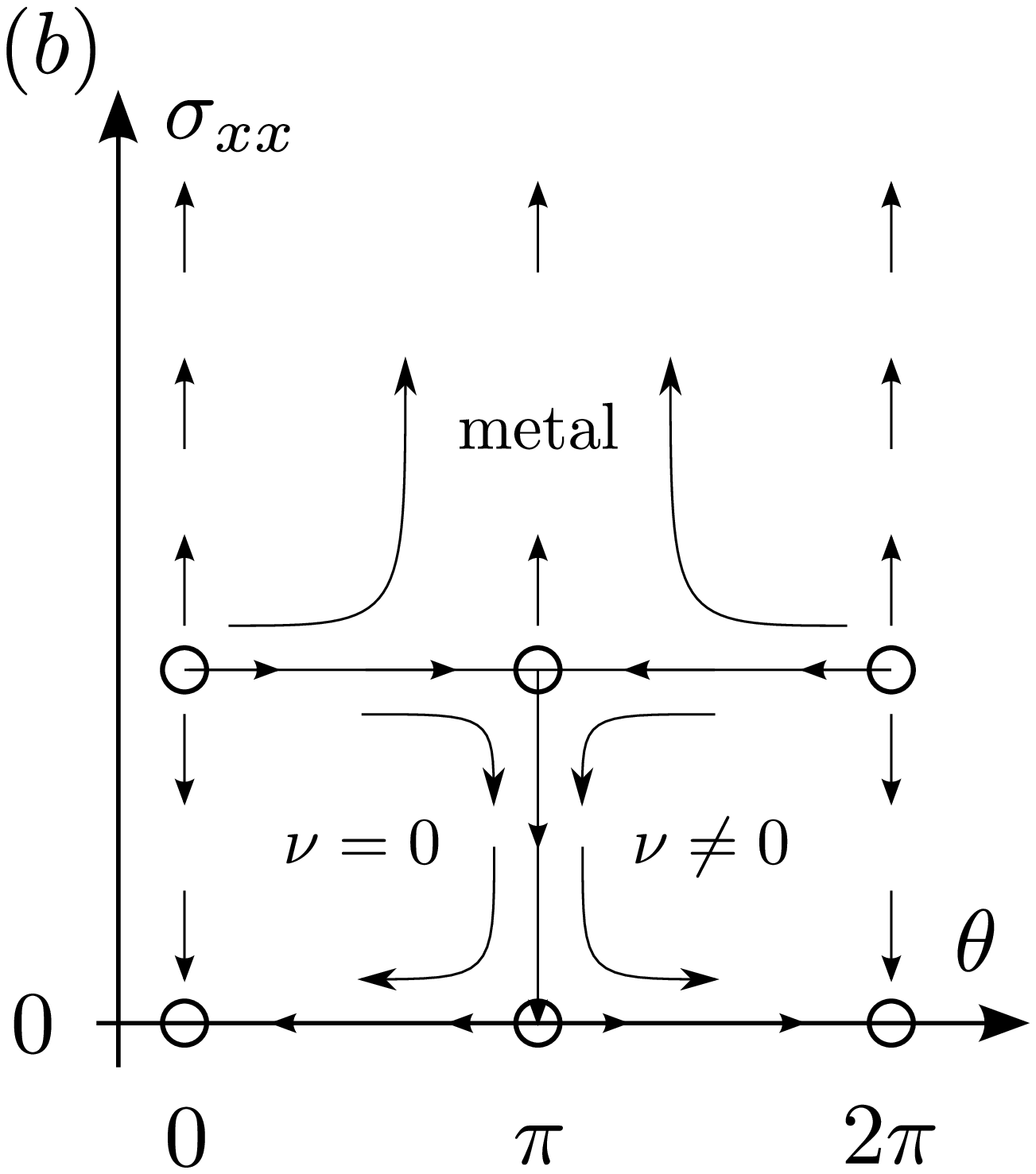}
\caption{
Possible two-parameter scaling flow diagrams of 
symmetry class CI and DIII.
        } 
\label{fig:ci-diii}
\end{figure}

We first discuss the phase diagram for class CI and DIII.
(Symmetry class AIII has a complication due to the special nature of 
its beta function in ($2+\epsilon$) dimensions -- see below).
The beta functions for these NL$\sigma$Ms in 3d when  $\theta=0$
are expected to have 
a zero at finite $\sigma_{xx}$,
separating stable metallic and 
Anderson insulating phases.
For large $\sigma_{xx}$, 
the RG flow for arbitrary $\theta\neq 0$ 
is expected to be similar to 
the RG flow at $\theta=0$,
while the flow for $\theta\neq 0$ and for small $\sigma_{xx}$
needs to be determined.
In terms of the massive Dirac model with disorder,
on the other hand, 
$\sigma_{xx}$ is small around the clean Dirac fermion point
(the dc conductivity of the clean 3d Dirac fermion is zero)
and $\theta\sim \pi$.
It is thus tempting to ``identify'' 
the point $(\sigma_{xx}, \theta)=(0, \pi)$ in the NL$\sigma$M
phase diagram as the clean Dirac fixed point,
or, at least, to use the Dirac model (i.e., the chiral Gross-Neveu model) 
to guess the RG flow of the NL$\sigma$M.

The fermion mass term $\bar{\psi}\gamma_5\psi$
in the chiral Gross-Neveu model 
is a relevant perturbation to the clean Dirac fixed point.
Also, depending on the sign of the mass,
one can drive the clean Dirac fermion either into 
the topological ($\nu\neq 0$)
or trivial ($\nu=0$) insulating phase.
As the fermion mass corresponds to the theta term in the NL$\sigma$M, 
this would suggest the deviation of the theta angle from $\theta=\pi$
is relevant at $\sigma_{xx}=0$.
On the other hand, disorder at the clean Dirac fixed point is irrelevant
by power-counting. 
Furthermore,
the existence of the UV fixed point (the Gross-Neveu fixed point)
in the $1/N_f$-expansion of the chiral Gross-Neveu model
would suggest the existence of a corresponding critical point at $\theta=\pi$
in the NL$\sigma$M
(Fig.\ \ref{fig:ci-diii})
separating  a metallic phase and an insulating phase
(which is adiabatically connected the clean 3d Dirac fermion with 
$\sigma_{xx}=0$). 

While we expect the deviation of the theta angle from $\theta=\pi$
is relevant at $\sigma_{xx}=0$ (i.e., at the clean Dirac fixed point),
there is an ambiguity on the RG flow around 
the putative critical point at $\theta=\pi$ and $\sigma_{xx}\neq 0$. 
If it is relevant, as it is at the clean Dirac point
$(\sigma_{xx},\theta)=(0,\pi)$, 
the flow looks like Fig.\ \ref{fig:ci-diii}-(a). 
An alternative scenario where the theta term is irrelevant 
at the Gross-Neveu critical point is depicted in Fig.\ \ref{fig:ci-diii}-(b). 
In these scenarios, 
an implicit assumption we have made is that there are no other fixed points.
Also, we have not considered more exotic possibilities, 
such as the case where $\theta$ is exactly marginal at the Gross-Neveu fixed point.
Which one of the two scenarios is realized may be speculated,
again, from the large $N_f$ approach, 
by computing the anomalous dimension of the
$\gamma_5$ fermion mass;
See Eqs.\
(\ref{mass dim AIII}),
(\ref{gamma5 mass dim AIII}),
and
(\ref{mass dim DIII}) for relevant calculation.
We should, of course, bear in mind that
the structure of the phase diagram would be quite different
for large $N_f$ and for small $N_f$ (e.g., $N_f\to 1$). 
In other words,
with the inclusion of the $N_f$ flavor of Dirac fermions,
the model describes, 
in the absence of disorder, a transition between 
trivial and topological insulators with $\nu=N_f$.
This is a highly fine tuned critical point,
and might be quite different from the critical point
that may exist and be described in the NL$\sigma$M.

Symmetry class AIII is special in that
the beta function in $(2+\epsilon)$-expansion is always positive;
Thus, at $\theta=0$, there is no metal insulator transition,
but we always have a metal.
However, as the large-$N_f$ expansion (which corresponds to $\theta\neq 0$) 
predicts 
tuning $\theta\neq 0$ seems to generate a metal-insulator transition. 
Because of the positivity of the beta function in $(2+\epsilon)$ dimensions, 
it appears that
constructing a reasonable conjecture on the two-parameter scaling 
phase diagram is more complicated.

\section{discussion}

Topological insulators and superconductors are
known to be an inherently holographic system;
The non-trivial bulk topology 
[as encoded by a non-vanishing bulk topological invariant 
such as the Chern (TKNN)\cite{TKNN82} integer or the $\mathbb{Z}_2$ invariant 
in $\mathbb{Z}_2$ topological insulators]
has a crucial impact on the Anderson localization problem
at boundaries of the system;
in fact, Anderson localization is not allowed at the system's boundaries 
because of topological reasons. 
\cite{Nomura07,Z2anomaly07,Ostrovsky07}
Such bulk-boundary correspondence served 
as a guiding principle of classification
of topological insulators and superconductors. 
In this paper, we have investigated Anderson localization 
problem {\it in the bulk} of topological insulators and superconductors
in three dimensions;
we have explored implications of bulk topology on physics 
of Anderson localization in the bulk.

In the QHE, the transition between topological phases with
different Chern integers
is governed by a disorder-dominated quantum critical point
(the plateau transition).
The fact that delocalized (extended) states 
are possible at the plateau transition
is a direct manifestation of a topological character 
of the system, which is not allowed otherwise 
in two dimensions with broken time-reversal symmetry. 
In the NL$\sigma$M description of the QHE, 
the plateau transition is encoded in the Pruisken term
(which also points the existence of edge modes).
We can draw a rather close analogy 
between time-reversal symmetric topological superconductors 
in three dimensions and the 2d QHE,
both of which are characterized by an integer topological invariant
in the bulk.
We have given a field theory description of 
metal-insulator transition in terms of the Dirac model
of topological superconductors 
by employing the large-$N_f$ expansion,
and also in terms of the NL$\sigma$Ms. 
Although the presence of topological terms 
in the NL$\sigma$Ms is indicative,
it is an open question if the phase diagram is characterized in terms of 
two parameter scaling, as in the QHE.
It is desirable to explore further
the structure of the phase diagram and
critical behaviors in 3d topological
superconductor systems in terms of, say,  
numerics, preferably in the presence of
the non-zero theta angle in the language of the NL$\sigma$M.
This would require more microscopic and physical understanding of 
the theta angle; 
e.g., how one should interpret and measure
the theta term in a metallic phase. 
\cite{Barkeshli2011,Bergman2011}

Our discussion for symmetry class DIII can be relevant, 
for example, for $^3$He in aerogel,
\cite{Porto1995,Halperin2003}
although we need to include pair breaking effects.
And also 
Cu-intercalated topological insulators, Cu$_x$Bi$_2$Se$_3$, 
\cite{Hor2010,Wray2010,Sasaki2011,FuBerg2010}
and
many heavy fermion non-centrosymmetric superconductors
with antisymmetric spin-orbit coupling. 
\cite{noncentro10}
Experimental probes for 
detecting these systems are, for example, 
thermal transport,
and
spin transport
when spin or part of spin quantum number is conserved.
\cite{Ryu2011,Wang2011}
See, for example, 
for the anomalous thermal Hall effect by magnons,
Ref.\ \onlinecite{Onose2010}.

In the 3d $\mathbb{Z}_2$ topological insulator in symplectic symmetry class,
a closer analogy would be the 2d quantum spin Hall effect.
We have numerically investigated the structure of the phase 
diagram by the transfer matrix method, and developed 
the large-$N_f$ field theory approach to a metal-insulator transition.
It should be noted, however, 
the large $N_f$ expansion does not distinguish 
the case of odd $N_f$, for which the one of the insulating
states in the phase diagram Fig.\ \ref{fig: phase dirgram} is
a topologically non-trivial, $\mathbb{Z}_2$ topological insulator,
and the case of even $N_f$,
for which the one of the insulating
states in the phase diagram Fig.\ \ref{fig: phase dirgram} is
a topologically trivial. 
While we have imposed the global $\mathrm{O}(N_f)$ symmetry in the large
$N_f$ expansion,
perhaps a distinction between even and odd $N_f$ would emerge 
once we study the stability of 
the UV fixed point (the Gross-Neveu fixed point)
that can be reached by the large $N_f$ expansion; 
for example, 
for even $N_f$, 
the Gross-Neveu fixed point is unstable against 
$\mathrm{O}(N_f)$ symmetry breaking (and other) perturbations,
and the metal-insulator transition is always 
controlled by the conventional critical point
of symplectic symmetry class.
In terms of the NL$\sigma$M description,
there is no topological term of any kind in the bulk
in symmetry class AII,
and hence the transition between the metal and 
insulating phases is expected in the conventional universality class
of symplectic symmetry class, 
irrespective of the number of flavors.
This situation is similar to 
the $\mathbb{Z}_2$ topological insulators in
class AII in two dimensions\cite{QSHnum2,QSHnum3}.

While we have discussed the bulk critical behavior,
it would be interesting to study boundary critical behaviors;
even when a transition 
between a trivial insulator to a metal
and 
one between a topological insulator to a metal
is in the same universality class, 
at a surface we might observe different critical behavior.
This was indeed the case in the 2d symplectic class.
\cite{obuse2008}

Finally, we conclude with some remarks on 
other symmetry classes in three dimensions.

--
For symmetry class CII, 
a $\mathbb{Z}_2$ topological insulator can be realized in the bulk.
\cite{Hosur2010}
While we did not discuss symmetry class CII, 
a similar field theory analysis can be developed. 

-- 
The large-$N_f$ technique developed here is also applicable 
to the Weyl semimetal in symmetry class A,
which are realized
in the A phase of $^3$He
and Pyrochlore Iridates.
\cite{Wan2011, Volovikflatband2011}

-- 
We mention two other symmetry classes 
for which the corresponding NL$\sigma$M has
a topological term of $\mathbb{Z}_2$ type:
in symmetry class C and CII,
$\mathbb{Z}_2$ insulators (superconductors) 
can be realized in four dimensions.
Thus, the 3d $\mathbb{Z}_2$ topological term in the corresponding
NL$\sigma$M describes the surface of 4d insulator,
but not any isolated 3d system.

\appendix

\acknowledgments

The authors acknowledge helpful interactions with
Hideaki Obuse, and Ai Yamakage.
S.R. thanks RIKEN for hospitality.
This research is granted by the Japan Society for the Promotion of Science (JSPS) through the "Funding Program for World-Leading Innovative R{\&}D on Science and Technology (FIRST Program)," initiated by the Council for Science and Technology Policy (CSTP).

\appendix
\section{transfer matrix}
\label{transfer matrix}

\paragraph{transfer matrix}

In this appendix, we derive the transfer matrix for four-component 
3d Dirac fermions.
The Dirac Hamiltonian is written in the single-particle representation as
\begin{align}
H
&=
-\sum_{n,{\bf r}}
\left[
|{\bf r},n+1,\sigma\rangle 
{u}^z_{-,\sigma\sigma'}
\langle {\bf r},n,\sigma'|
\right.
\nonumber \\
&\quad
\left.
\qquad \quad 
+
|{\bf r},n,\sigma\rangle 
{u}^z_{+,\sigma\sigma'}
\langle {\bf r},n+1,\sigma|
\right] 
\nonumber \\
&\quad
-
\sum_{n,{\bf r}}
\sum_{{\mu=1,2}}
\left[
|{\bf r+e}_{\mu},n,\sigma\rangle 
{u}^{\mu}_{-,\sigma\sigma'}
\langle {\bf r},n,\sigma'|
\right.
\nonumber \\
&\quad
\left. 
\qquad \quad 
+
|{\bf r},n,\sigma\rangle 
{u}^{\mu}_{+,\sigma\sigma'}
\langle {\bf r+e}_{\mu},n,\sigma'|
\right]
  \nonumber \\
&\quad 
+\sum_{n,{\bf r}}
|{\bf r},n,\sigma\rangle
[(m+3)\beta+\varepsilon_{{\bf r}n}]_{\sigma\sigma'}
\langle {\bf r},n,\sigma'|, 
\end{align}
where we have introduced 
\begin{align}
{u}^{\mu}_{\pm}
=
\frac{1}{2}
\left(
\beta \pm it\alpha_{\mu}
\right).
\end{align}
The eigen wave function with energy $E$ of this Hamiltonian satisfies
\begin{align}
0&=\langle {\bf r}n\sigma|(H-E)|\psi \rangle \nonumber \\
 &=
- 
{u}^{z}_{-,\sigma\sigma'}
\psi_{\sigma'}({\bf r},n-1)
- 
{u}^{z}_{+,\sigma\sigma'}
\psi_{\sigma'}({\bf r},n+1)
\nonumber \\
&\quad
+[\mathcal{H}^{(n)}-E]_{\sigma\sigma'}\psi_{\sigma'}({\bf r},n),
\end{align}
which leads to 
\begin{align}
 \psi_{n+1}
&=
2(\beta+it\alpha_z)^{-1}[\mathcal{H}^{(n)}-E]\psi_n
\nonumber \\
&\quad
 -(\beta+i\alpha_z)^{-1}(\beta-i\alpha_z)\psi_{n-1}. 
\end{align}
Note that 
$(\beta+it\alpha_z)^{-1}=(\beta+it\alpha_z)/(1-t^2)$, 
and Eq.\ (\ref{psi_n}) is obtained.
The transfer matrix $M^{(n)}$ is defined as
\begin{align}
\begin{pmatrix}
\psi_{n+1} \\
\psi_{n}
\end{pmatrix}=M^{(n)}
\begin{pmatrix}
\psi_{n} \\
\psi_{n-1}
\end{pmatrix},\quad
  M^{(n)}=
\begin{pmatrix}
h^{(n)} & v \\
1 & 0
\end{pmatrix}, 
\end{align}
where 
\begin{align}
h^{(n)}
&\equiv 
\frac{-4}{t^2-1}
{u}^z_+ 
[\mathcal{H}^{(n)}-E]\nonumber \\
&=
\sum_{n,{\bf r},\mu=1,2}
|{\bf r+e}_{\mu},n,\sigma\rangle 
\left[
\frac{4 {u}^z_{+} {u}^z_{-}}{t^2-1}
\right]_{\sigma\sigma'}
\langle {\bf r},n,\sigma'|
\nonumber \\
&\quad   +
\sum_{n,{\bf r},\mu=1,2}
| {\bf r},n,\sigma\rangle 
\left[
\frac{4{u}^{z}_{+} {u}^{\mu}_{+}}{t^2-1}
\right]_{\sigma\sigma'}
\langle {\bf r+e}_{\mu},n,\sigma'|
\nonumber \\
&\quad
+
\sum_{n,{\bf r}}
| {\bf r},n,\sigma\rangle 
\frac{-4}{t^2-1} 
\Big[(m+3)
{u}^{z}_{+}\beta 
\nonumber \\
&\quad
\qquad \qquad \qquad 
+
(\varepsilon_{{\bf r}n}-E)
{u}^{z}_{+} 
\Big]_{\sigma\sigma'}
\langle {\bf r},n,\sigma'|. 
\end{align}

\section{large-$N_f$ expansion for 
three-dimensional topological 
superconductors}
\label{sec:
large-$N_f$ expansion for 
three-dimensional topological 
superconductors}

In this Appendix, we illustrate the large-$N_f$ calculations  
for 3d time-reversal symmetric topological superconductors 
in symmetry class AIII, DIII, and CI.
We start, in Subsec.\ \ref{fermionic replica field theories},
by reviewing fermionic replica methods 
for symmetry class AIII, DIII, and CI.
The formalism in Subsec.\ \ref{fermionic replica field theories},
is general and applies to any systems in these symmetry classes
in any dimensions other than 3d Dirac representatives
of topological superconductors discussed in the main text. 
In Subsec.\ \ref{auxiliary matrix fields}, 
we introduce auxiliary matrix fields (Hubbard-Stratonovich fields)
to decouple four fermion interactions induced by
quenched disorder averaging,
which is useful to develop the large-$N_f$ expansion,
and also to derive the NL$\sigma$M inside metallic phases. 

\subsection{fermionic replica field theories}
\label{fermionic replica field theories}

For symmetry classes AIII, DIII, and CI, 
the Hamiltonians anticommute with a unitary matrix 
$\Gamma_z$ with equal number of eigenvalues $+1$ and $-1$,
$\{\mathcal{H}, \Gamma_z \}=0$, 
thereby they can be brought into an block off-diagonal form, 
\begin{align}
\mathcal{H}
&=
\left(
\begin{array}{cc}
0 & D \\
D^{\dag} & 0
\end{array}
\right), 
\end{align}
in the basis where $\Gamma_z$ is written diagonally as
\begin{align}
\Gamma_z 
&=
\left(
\begin{array}{cc}
1 & 0 \\
0 & -1
\end{array}
\right).
\end{align}
For class AIII, the Hamiltonians are not subjected to
any further constraint. On the other hand, 
for classes DIII and CI, the off-diagonal block $D$ 
satisfies 
\begin{align}
D^T &= - D
\quad
\mbox{for class DIII},
\nonumber \\
D^T &= + D
\quad
\mbox{for class CI}.
\end{align}

In the following, we will assume that the Hamiltonian consists
of the disorder-free ($\mathcal{H}_0$) 
and disordered ($\mathcal{V}$) parts, 
\begin{align}
\mathcal{H}= \mathcal{H}_0 + \mathcal{V},
\quad
D = D_0 + D_I(x),
\end{align}
where the random potential
$\mathcal{V}$ is assumed to be white-noise and to be maximally random
according to the distribution
\begin{align}
\propto 
\exp\left[
-\mathrm{tr}\,\big(D^{\ }_ID_I^{\dag}\big)/(2g)
\right],
\end{align}
with $g$ characterizing the strength of disorder. 

In order to study Anderson localization physics,
we are after the properties of the single particle 
Green's function
$G(z)= (z-\mathcal{H})^{-1}$
at zero energy, $\mathrm{Re}\,z =0$.
Since ``chiral'' symmetry $\{\mathcal{H},\Gamma_z\}=0$
relates retarded and advanced 
Green's functions at zero energy, 
we can focus either one of them,
the retarded Green's function, say,
$G(z={i}0^+)= ({i}0^+-\mathcal{H})^{-1}$,
where $0^+>0$ is positive infinitesimal.
The generating function for the retarded Green function
can be expressed in terms of fermionic functional integral,
$Z=
\int \mathcal{D}[f^{\dag},f]
\exp(-\int d^3x\, \mathcal{L} )$,
\begin{align}
\mathcal{L}
&= 
f^{\dag}
\left(
{i}\mathcal{H} + 0^+ 
\right)
f, 
\end{align}
where $f^{\dag}$ and $f$ are independent fermionic path integral variables.

To perform the $1/N_f$ expansion, 
we generalize the above functional integral with single flavor 
to
$Z=
\int \mathcal{D}[f^{\dag}_{\iota},f^{\ }_{\iota}]
\exp(-\int d^3x\, \mathcal{L} )$,
\begin{align}
\mathcal{L}
&= 
\sum^{N_f}_{\iota,\kappa=1}
f^{\dag}_{\iota}
\left[
\left(
{i}\mathcal{H}_0+0^+
\right)
\delta_{\iota\kappa} 
+
{i}\mathcal{V}_{\iota\kappa}
\right]
f_{\kappa},
\end{align}
where the disorder part 
\begin{align}
\mathcal{V}_{\iota\kappa}
=
\left(
\begin{array}{cc}
0 & D_{I,\iota\kappa} \\
D^{\dag}_{I,\iota\kappa} & 0
\end{array}
\right)
\end{align}
is a suitable generalization of $\mathcal{H}_I$
which couples all flavors equally
according to the disorder distribution
$\propto 
\exp\big[
-\mathrm{tr}\,(D^{\ }_{I,\iota\kappa}D^{\dag}_{I,\kappa\iota})/(2g)
\big]$.

\subsubsection{class AIII}

Quenched averaging over disorder $D_I$ can be done 
by introducing replicas
$\{f^{\dag}_{a\iota}, f^{\dag}_{a\iota}\}_{a=1,\ldots, N_r}$ 
which leads to the replicated Lagrangian 
\begin{align}
&\quad \mathcal{L}_{\mbox{\begin{tiny}AIII\end{tiny}}}
= 
\psi^{\dag}_{a \iota}
({i}\mathcal{H}_0+0^+ )
\psi^{\ }_{a \iota}
\\
&\quad
\qquad 
-
\frac{g}{2}
\big(
\psi^{\dag}_{a\iota}  \psi^{\ }_{b \iota}
\psi^{\dag}_{b\kappa}  \psi^{\ }_{a\kappa}
-
\psi^{\dag}_{a\iota} \Gamma_z  \psi^{\ }_{b \iota}
\psi^{\dag}_{b\kappa} \Gamma_z \psi^{\ }_{a\kappa}
\big),
\nonumber 
\end{align}
where we have renamed 
$f^{\dag},f
\to 
\psi^{\dag},\psi$,
and repeated replica and flavor indices 
are implicitly summed. 

When we deal with the relativistic model
(massive Dirac model) 
of the topological superconductor, 
it will prove convenient to introduce 
\begin{align}
\bar{\psi} := \psi^{\dag} {i} \Gamma_z. 
\end{align}
The Lagrangian is then given as
\begin{align}
\label{chiral GN for AIII}
&\quad \mathcal{L}_{\mbox{\begin{tiny}AIII\end{tiny}}}
= 
\bar{\psi}^{\ }_{a \iota}
(\Gamma_z \mathcal{H}_0 -{i} \Gamma_z 0^+ )
\psi^{\ }_{a \iota}
\\
&\quad
\quad 
+
\frac{g}{2}
\big(
\bar{\psi}^{\ }_{a\iota} \Gamma_z \psi^{\ }_{b \iota}
\bar{\psi}^{\ }_{b\kappa}  \Gamma_z \psi^{\ }_{a\kappa}
-
\bar{\psi}^{\ }_{a\iota} \psi^{\ }_{b \iota}
\bar{\psi}^{\ }_{b\kappa} \psi^{\ }_{a\kappa}
\big). 
\nonumber 
\end{align}
This is the chiral Gross-Neveu model
with $\mathrm{U}(N_r)\times \mathrm{U}(N_r)$ internal symmetry.

\subsubsection{class DIII}

For class DIII and (and class CI, which will be discussed later), 
it is advantageous to double the number of components 
of fermionic fields
to account for the symmetry of the Hamiltonian,
$D=\pm D^T$.
For class DIII, 
we first note that
\begin{align}
\int d^3 x\, f^{\dag}_{A}
D
f^{\ }_{B}
&=
+
\int d^3 x\,  f^{T}_{B}
D
f^{* }_{A},
\nonumber \\
\int d^3 x\,  f^{\dag}_{B}
D^{\dag}
f^{\ }_{A}
&=
+
\int d^3 x\,  f^{T}_{A}
D^{\dag}
f^{* }_{B},
\end{align}
where 
$f=\left(f^{\ }_A, f^{\ }_B\right)^T$
in the chiral (or ``sublattice'') grading. 
Introducing the ``charge-conjugation'' space by
\begin{align}
\chi^{\ }_+
=
\frac{1}{\sqrt{2}}
\left(
\begin{array}{c}
f^{\ }_B \\
f^*_A
\end{array}
\right),
\quad
\chi^{\ }_-
=
\frac{1}{\sqrt{2}}
\left(
\begin{array}{c}
f^{\ }_A \\
f^*_B
\end{array}
\right),
\end{align}
the Lagrangian can be written as
\begin{align}
\mathcal{L}_{\mbox{\begin{tiny}DIII\end{tiny}}}
&=
\chi^{T}_+ {i}\tau_x \otimes D \chi^{\ }_+
+
\chi^{T}_- {i}\tau_x \otimes D^{\dag} \chi^{\ }_-
\nonumber \\
&\quad
+
\frac{0^+}{2}
\left(
\chi^{T}_+ \tau_x \tau_z \chi^{\ }_-
+
\chi^{T}_- \tau_x \tau_z \chi^{\ }_+
\right),
\end{align}
where the Pauli matrices $\tau_{0,x,y,z}$ acts on 
the ``charge-conjugation'' space.
Further redefining the fermionic fields as
\begin{align}
&\quad
\eta^{\ }_+ := \tau_{zy} \chi^{\ }_+,
\quad 
\eta^{\ }_- := \tau_{zy} \chi^{\ }_-,
\\
\mbox{where}\quad 
&\quad
\tau_{zy} = \left(\tau_z +\tau_y\right)/\sqrt{2},
\quad
\tau_{zy}^T \tau^{\ }_{zy}
=
-{i}\tau_x 
,
\nonumber 
\end{align}
the Lagrangian can be written as
\begin{align}
\mathcal{L}_{\mbox{\begin{tiny}DIII\end{tiny}}}
&=
{i}
\left(
{i}\eta^T_+ \tau_0\otimes D \eta^{\ } _+
+
{i}\eta^T_- \tau_0\otimes D^{\dag} \eta^{\ }_-
\right)
\nonumber \\
&\quad
+
\frac{0^+}{2}
\left(
{i}
\eta^T_+ \tau_y  \eta_-
+
{i}
\eta^T_- \tau_y  \eta_+
\right),
\end{align}
where we noted 
$\tau_{zy} \tau_z \tau_{zy} = \tau_y$.

At this stage, 
we notice that 
$f^{\dag} \mathcal{H} f$
is invariant under 
$\mathrm{O}(2)\times \mathrm{O}(2)$
transformations,
\begin{align}
\eta^{\ }_+ \to O^{\ }_+ \eta^{\ }_+,
\quad
\eta^{\ }_- \to O^{\ }_- \eta^{\ }_-,
\end{align}
where $O_{\pm}\in \mathrm{O}(2)$.
On the other hand, 
the ``smearing term'' $f^{\dag} 0^+ f$
breaks it down to $\mathrm{O}(2)$.
Indeed, in order for $O_+$ and $O_-$ to leave the smearing
term invariant,
they cannot be independent and must be ``locked'' by
$O^{\ }_-= -{i}\tau_y O^{\ }_+ {i}\tau_y$,
where ${i}\tau_y \in \mathrm{O}(2)$.
With replicas,
the $\mathrm{O}(2)\times \mathrm{O}(2)$
symmetry is promoted to
$\mathrm{O}(2N_r )\times \mathrm{O}(2N_r)$
symmetry which is broken down to
$\mathrm{O}(2N_r )$
by the smearing (see below).

Introducing the notation,
\begin{align}
\psi^{\dag}
=
{i}
\left(
\begin{array}{cc}
\eta^T_+, & \eta^T_-
\end{array}
\right),
\quad 
\psi
=
\left(
\begin{array}{c}
\eta_- \\
\eta_+
\end{array}
\right),
\quad
\end{align}
the Lagrangian is written as 
\begin{align}
\mathcal{L}_{\mbox{\begin{tiny}DIII\end{tiny}}}
&=
\psi^{\dag} \left(
{i}\tau_0 \otimes \mathcal{H}_0
+
0^+ \tau_y \otimes I
\right)
\psi, 
\end{align}
where we have rescaled $0^+/2 \to 0^+$. 
Observe that $\psi^{\dag}$ and $\psi$ satisfy 
the ``Majorana'' condition 
\begin{align}
\psi^T {i} \Gamma_x = \psi^{\dag}. 
\end{align}

With $N_f$ flavors and $N_r$ replicas, 
the Lagrangian after quenched disorder averaging can be 
cast into the form similar to the one for class AIII, 
\begin{align}
\mathcal{L}_{\mbox{\begin{tiny}DIII\end{tiny}}}
&=
\psi^{\dag}_{a\iota}
\left[
{i}\mathcal{H}_0\delta_{ab}+ 
0^+ (\tau_y\otimes I_{N_r})_{ab}
 \right] 
\psi^{\ }_{b\iota}
 \\
&\quad
-
\frac{g}{2}
\big(
\psi^{\dag}_{a\iota}  \psi^{\ }_{b\iota}
\psi^{\dag}_{b\kappa}  \psi^{\ }_{a\kappa}
-
\psi^{\dag}_{a\iota} \Gamma_z \psi^{\ }_{b\iota}
\psi^{\dag}_{b\kappa} \Gamma_z \psi^{\ }_{a\kappa}
\big),
\nonumber 
\end{align}
where
$a,b$ represents
the combined charge conjugation and
replica index
running from 1 to $2N_r$.

If we introduce 
\begin{align}
\bar{\psi} := \psi^{\dag} {i} \Gamma_z,
\quad 
\psi^T{i} \Gamma_y  = \bar{\psi},
\end{align}
the Lagrangian is then given as
\begin{align}
\label{chiral GN for DIII}
&\quad \mathcal{L}_{\mbox{\begin{tiny}DIII\end{tiny}}}
= 
\bar{\psi}^{\ }_{a \iota}
\left[
\Gamma_z \mathcal{H}_0 \delta_{ab}
+
0^+ \Gamma_z 
(-{i} \tau_y)_{ab} 
\right] 
\psi^{\ }_{b \iota}
\\
&\quad
\quad 
+
\frac{g}{2}
\big(
\bar{\psi}^{\ }_{a\iota} \Gamma_z \psi^{\ }_{b \iota}
\bar{\psi}^{\ }_{b\kappa}  \Gamma_z \psi^{\ }_{a\kappa}
-
\bar{\psi}^{\ }_{a\iota} \psi^{\ }_{b \iota}
\bar{\psi}^{\ }_{b\kappa} \psi^{\ }_{a\kappa}
\big). 
\nonumber 
\end{align}
This is the chiral Gross-Neveu model
with $\mathrm{O}(2N_r)\times \mathrm{O}(2N_r)$ internal symmetry.

\subsubsection{class CI}

To account for the class CI symmetry $D^T = + D$, 
we note that
\begin{align}
\int d^3x\,
f^{\dag}_{A}
D
f^{\ }_{B}
&=
-
\int d^3x\,
f^{T}_{B}
D
f^{* }_{A},
\nonumber \\
\int d^3x\,
f^{\dag}_{B}
D^{\dag}
f^{\ }_{A}
&=
-
\int d^3x\,
f^{T}_{A}
D^{\dag}
f^{* }_{B}, 
\end{align}
and introduce
\begin{align}
 \chi^{\ }_+
 =
 \frac{1}{\sqrt{2}}
 \left(
 \begin{array}{c}
 f^{\ }_B \\
 -f^*_A
 \end{array}
 \right),
 \quad
 \chi^{\ }_-
 =
 \frac{1}{\sqrt{2}}
 \left(
 \begin{array}{c}
 f^{\ }_A \\
 -f^*_B
 \end{array}
 \right).
\end{align}
The Lagrangian is then written as
\begin{align}
\mathcal{L}_{\mbox{\begin{tiny}CI\end{tiny}}}
&=
\chi^{T}_+ \tau_y \otimes D \chi^{\ }_+
+
\chi^{T}_- \tau_y \otimes D^{\dag} \chi^{\ }_-
\nonumber \\
&\quad
+
\frac{0^+}{2}
\left[
\chi^{T}_+  (-{i}\tau_y) \chi^{\ }_-
+
\chi^{T}_-  (-{i}\tau_y)\chi^{\ }_+
\right].
\end{align}

At this stage, one notices that 
$f^{\dag} \mathcal{H}f$ is invariant under 
transformations 
\begin{align}
\chi^{\ }_+ \to O^{\ }_+ \chi^{\ }_+,
\quad
\chi^{\ }_- \to O^{\ }_- \chi^{\ }_-,
\end{align}
where $O^{\ }_{\pm}$ satisfy
\begin{align}
O^T_{\pm} {i}\tau_y O^{\ }_{\pm } = {i}\tau_y.
\end{align}
I.e., $O^{\ }_{\pm} \in \mathrm{Sp}(1)$. 
The system is invariant under 
continuous $\mathrm{Sp}(1)\times \mathrm{Sp}(1)$ rotations.
The smearing term breaks this symmetry down to 
its diagonal subgroup;
in order for $O_{\pm}$ to leave the smearing
term invariant,
\begin{align}
O^T_+ {i}\tau_y O_- = {i}\tau_y
&\Rightarrow&
O_+ = O_-.
\end{align}
With replicas,
the $\mathrm{Sp}(1)\times \mathrm{Sp}(1)$
symmetry is promoted to
$\mathrm{Sp}(N_r )\times \mathrm{Sp}(N_r)$
symmetry which is broken down to
$\mathrm{Sp}(N_r )$
by the smearing term.

Let us now introduce 
\begin{equation}
\psi^{\dag}
=
 \left(
\begin{array}{cc}
\chi^{T}_+ (-{i}\tau_y), & \chi^{T}_- (-{i}\tau_y)
\end{array}
\right),
\quad 
\psi
=
\left(
\begin{array}{c}
\chi^{\ }_- \\
\chi^{\ }_+
\end{array}
\right). 
\end{equation} 
They satisfy the ``Majorana'' condition 
\begin{align}
\psi^{\dag}
=\psi^T \Gamma_x (-{i}\tau_y). 
\end{align}
The Lagrangian is written as
\begin{align}
\mathcal{L}_{\mbox{\begin{tiny}CI\end{tiny}}}
&=
\psi^{\dag}
\left(
\tau_0 \otimes {i}\mathcal{H}_0
+
0^+ 
\right)
\psi
=
\sum_{\tau=\pm}
\psi^{\dag}_\tau
\left(
{i}\mathcal{H}_0
+
0^+ 
\right)
\psi_{\tau}, 
\end{align}
where 
in the second line we made the particle-hole indices explicit,
and we have rescaled $0^+/2 \to 0^+$.

Introducing $N_f$ flavors
and $N_r$ replicas and then performing the 
quench disorder averaging, 
\begin{align}
&\quad
\mathcal{L}_{\mbox{\begin{tiny}CI\end{tiny}}}
=
\psi^{\dag}_{a\sigma \iota}
\left(
{i}
\mathcal{H}_0 
+
0^+  
\right)
\delta_{\sigma \tau} \delta_{ab}
\psi^{\ }_{b\tau \iota}
\\
&\quad
\quad 
-
\frac{g}{2}
\big(
\psi^{\dag}_{b\tau\iota} \psi^{\ }_{a\sigma \iota}
\psi^{\dag}_{a\sigma\kappa} \psi^{\ }_{b\tau\kappa}
-
\psi^{\dag}_{b\tau\iota} c_z \psi^{\ }_{a\sigma\iota}
\psi^{\dag}_{a\sigma\kappa} c_z \psi^{\ }_{b\tau\kappa}
\big),
\nonumber
\end{align}
where $a,b=1,\ldots, N_r$,
$\iota,\kappa=1,\ldots, N_f$,
and $\sigma,\tau=\pm$. 

If we introduce 
\begin{align}
\bar{\psi} := \psi^{\dag} {i} \Gamma_z,
\quad 
\psi^T
({i}\tau_y)
\Gamma_y  = \bar{\psi},
\end{align}
the Lagrangian is then given as
\begin{align}
\label{chiral GN for CI}
&\quad \mathcal{L}_{\mbox{\begin{tiny}CI\end{tiny}}}
= 
\bar{\psi}^{\ }_{a \sigma \iota}
\left(
\Gamma_z \mathcal{H}_0
- 
0^+ {i} \Gamma_z 
\right) 
\psi^{\ }_{a \sigma \iota}
\\
&\quad
\quad 
+
\frac{g}{2}
\big(
\bar{\psi}^{\ }_{a \sigma\iota} 
\Gamma_z \psi^{\ }_{b \tau \iota}
\bar{\psi}^{\ }_{b \tau \kappa}  
\Gamma_z \psi^{\ }_{a \sigma\kappa}
-
\bar{\psi}^{\ }_{a\sigma \iota} 
\psi^{\ }_{b \tau \iota}
\bar{\psi}^{\ }_{b \tau\kappa} 
\psi^{\ }_{a \sigma \kappa}
\big). 
\nonumber 
\end{align}
This is the chiral Gross-Neveu model
with $\mathrm{Sp}(N_r)\times \mathrm{Sp}(N_r)$ internal symmetry.

\subsection{auxiliary matrix fields}
\label{auxiliary matrix fields}

The four-fermion interactions in the replica field theories 
can be decoupled by two auxiliary matrix fields which 
we call $W$ and $V$. 
Depending on symmetry class,
these matrix fields satisfy different constraints.

\subsubsection{class AIII}
For symmetry class AIII, 
the fermionic Lagrangian (\ref{chiral GN for AIII})
can be rewritten 
\begin{align}
\mathcal{L}_{\mbox{\begin{tiny}AIII\end{tiny}}}
&=
\bar{\psi}_{a,\iota}
(\Gamma_z\mathcal{H}_0 -{i}0^+ \Gamma_z)
\delta_{ab}
\psi_{b,\iota}
\nonumber \\
&\quad
+
\bar{\psi}_{a,\iota}
\left(
W_{ab}
+
{i} 
\Gamma_z V_{ab}
\right)
\psi_{b,\iota}
\nonumber \\
&\quad
+\frac{1}{2g}
\mathrm{tr}^{\ }_{N_r}\,\left(
W^2 + V^2
\right),
\end{align}
where
$W$ and $V$ are an $N_r\times N_r$ hermitian matrix. 

\subsubsection{class DIII}

For symmetry class DIII, 
the fermionic Lagrangian (\ref{chiral GN for DIII})
can be rewritten 
\begin{align}
\mathcal{L}_{\mbox{\begin{tiny}DIII\end{tiny}}}
&=
\bar{\psi}^{\ }_{a \iota}
\left[
\Gamma_z \mathcal{H}_0 \delta_{ab}
+
0^+ \Gamma_z 
(-{i} \tau_y \otimes I_{N_r})_{ab} 
\right] 
\psi^{\ }_{b \iota}
\nonumber \\
&\quad
+
\bar{\psi}^{\ }_{a\iota}
\left(
W_{ab}
+{i}
\Gamma_z V_{ab}
\right)
\psi^{\ }_{b\iota}
\nonumber \\
&\quad
+
\frac{1}{2g}
\mathrm{tr}^{\ }_{2N_r}\, 
\left(
W^2 + V^2
\right),
\end{align}
where 
$W$ is a $2N_r \times 2N_r$ real symmetric matrix,
while
$V$ is a $2N_r \times 2N_r$ pure imaginary antisymmetric matrix.

\subsubsection{class CI}

To decouple the four fermion terms 
in the fermionic Lagrangian (\ref{chiral GN for AIII})
for symmetry class CI,
we need to introduce two bosonic fields
$W_{a\tau, b\tau'}$ and
$V_{a\tau, b\tau'}$ 
where the first type of index $a,b$ 
runs over the replica grading, $a,b=1,\cdots, N_r$,
and the second type of index $\tau,\tau'$ 
runs over the particle-hole grading, $\tau,\tau'=\pm$. 
With these auxiliary fields, the
fermionic Lagrangian (\ref{chiral GN for AIII}) can be written as 
\begin{align}
\mathcal{L}_{\mbox{\begin{tiny}CI\end{tiny}}}
 &=
\bar{\psi}^{\ }_{a\sigma\iota}
(\Gamma_z\mathcal{H}_0 - {i} 0^+ \Gamma_{z})
\delta_{ab}\delta_{\sigma \tau}
\psi^{\ }_{b\tau\iota}
\nonumber \\
&\quad
+
\bar{\psi}^{\ }_{a \sigma \iota}
\left(
W_{a\sigma,b\tau}
+{i}\Gamma_z 
V_{a\sigma,b\tau}
\right)
\psi^{\ }_{b \tau \iota}
\nonumber \\
&\quad
+
\frac{1}{4g}
\mathrm{tr}_2 \, \left(
V^{\ }_{ba} V^{\ }_{ab}
+
W^{\ }_{ab} W^{\ }_{ba}
\right).
\end{align}
Here in the last line, 
the 2d trace is taken over the particle-hole space,
and $V_{ab}$ and $W_{ab}$,
for the fixed replica indices $a$ and $b$,
are viewed as a $2\times 2$ matrix. 
They are a real quartenion matrix,
i.e.,
they can be expanded as
$V_{ab}= A+B_x ({i}\tau_x) + B_y ({i}\tau_y)+B_z ({i}\tau_z)$
and
$W_{ab}= C+D_x ({i}\tau_x) + D_y ({i}\tau_y)+D_z ({i}\tau_z)$
with real coefficients $A, B_{x,y,z}, C, D_{x,y,z}$, 
and also satisfy
\begin{align}
W^{\ }_{ba}&= - (-{i}\tau_y) W^T_{ab} ({i}\tau_y)
\quad  
(\equiv -W^R_{ab}),
\nonumber \\
V^{\ }_{ba}&= 
+(-{i}\tau_y) V^T_{ab} ({i}\tau_y)
\quad
(\equiv +V^R_{ab}). 
\end{align}

\subsection{leading order analysis} 

To perform the $1/N_f$ expansion, 
we first compute the propagators
of $W$ and $V$ bosons dressed by
fermion bubbles. 
This is depicted diagrammatically 
in Fig.\ \ref{fig:diagrams}.

For symmetry class AIII,
propagators for $W$ and $V$ bosons are given by
\begin{align}
\langle
W_{ab}(q) W_{cd}(-q')
\rangle
&=
\langle
V_{ab}(q) V_{cd}(-q')
\rangle
\nonumber \\
&
=
\frac{4}{N_f D_{\gamma}} 
\mathcal{D}(q)
\delta_{q,q'}
\delta_{cb}
\delta_{da},
\end{align}
where $\mathcal{D}(q)$ is defined in 
Eg.\ (\ref{bosonic propagator}). 
Similarly, 
for class DIII, 
\begin{align}
\langle
W_{ab}(-q)
W_{cd}(q')
\rangle
&=
\frac{8}{N_f D_{\gamma}} 
\mathcal{D}(q)
\left(
\delta_{cb}\delta_{da}
+
\delta_{ca}\delta_{db}
\right)
\delta_{q,q'},
\nonumber \\
\langle
V_{ab}(-q)
V_{cd}(q')
\rangle
&=
\frac{8}{N_f D_{\gamma}} 
\mathcal{D}(q)
\left(
\delta_{cb}\delta_{da}
-
\delta_{ca}\delta_{db}
\right)
\delta_{q, q'}, 
\end{align}
for $a,b,c,d=1,\ldots, 2N_r$,
and for symmetry class CI, 
\begin{align}
&\quad
\langle
W_{a\sigma, b\tau}(-q)
W_{c\sigma', d\tau'}(q')
\rangle
=
\frac{8}{N_f D_{\gamma}} 
\mathcal{D}(q)
\delta_{q,q'}
\nonumber \\
&\quad
\quad \quad 
\times
\Big(
\delta_{cb}\delta_{da}\delta_{\sigma',\tau}\delta_{\tau',\sigma}
-\sigma \tau 
\delta_{ca}\delta_{db}\delta_{\sigma',-\sigma}\delta_{\tau',-\tau}
\Big), 
\nonumber \\
&\quad
\langle
V_{a\sigma, b\tau}(-q)
V_{c\sigma', d\tau'}(q')
\rangle
=
\frac{8}{N_f D_{\gamma}} 
\mathcal{D}(q)
\delta_{q,q'}
\nonumber \\
&\quad
\quad \quad 
\times 
\Big(
\delta_{cb}\delta_{da}
\delta_{\sigma' \tau}\delta_{\tau' \sigma}
+
\sigma \tau 
\delta_{ca}\delta_{db}\delta_{\sigma',-\sigma}\delta_{\tau',-\tau}
\Big). 
\end{align}

\section{non-linear sigma models and topological angle
for three-dimensional topological superconductors}
\label{non-linear sigma models and topological angle}

In this appendix, we will derive the
non-linear sigma model with topological term
for symmetry class AIII. 
A similar derivation should be possible for
symmetry class DIII and CI as well.

We start from the saddle point solution 
Eq.\ (\ref{saddle AIII}):
\begin{align}
(w,v) =(0, v)
\end{align}
where we have set $w=0$ since
such saddle points can be reached from 
$(0, v)$ by continuous $\mathrm{U}(N_r)\times \mathrm{U}(N_r)_A$ rotation.
I.e., 
from this saddle point, 
one can generate other saddle points as
$v\gamma^0 \exp (2{i}\gamma^0 \theta)$
where $\theta\in \mathrm{u}(N_r)$.
In other words, 
non-zero saddle point $v$
destroys the ``axial'' $\mathrm{U}(N_r)_A$ symmetry.
Such continuous rotation
is nothing but the Nambu-Goldstone mode which is the 
diffuson or Cooperon representing the diffusive motion of 
fermionic quasiparticles. 
The coupling of 
the Nambu-Goldstone mode to fermions are described by
\begin{align}
\mathcal{L} 
=
\bar{\psi}
\big(
\gamma^k \partial_k 
+m \gamma^5
+ v\gamma^0
e^{2{i}\gamma^0 \theta(r)}
\big)
\psi
+
\mathrm{const.}, 
\end{align}
where, with space-dependent $\theta$,
$\exp(2{i}\gamma^0 \theta(r))$
represents 
the Nambu-Goldstone modes, i.e., 
the fields in the NL$\sigma$M,
\begin{align}
e^{2{i}\gamma^0 \theta(r)}
=
\left(
\begin{array}{cc}
U & 0 \\
0 & U^{\dag}
\end{array}
\right),
\quad
U \in \mathrm{U}(N_r).
\end{align}

In the following, we will integrate over fermions
to derive the effective action of the Nambu-Goldstone mode.
We will do not include
the longitudinal fluctuations for a while,
but it is important as the integration over them
leads to the Gade term. 
\cite{Altland_Simons1999,Altland_Merkt2001}.
To prepare for the gradient expansion,
we will ``unwind'' $U$:
\begin{align}
\mathcal{L}
&= 
\bar{\psi}
\left(
\begin{array}{cc}
m_0 & \sigma_k  \partial_k + m_5 \\
U\left(- \sigma_k \partial_k + m_5  \right)U^{\dag} & -m_0 
\end{array}
\right)
\psi
\\
&= 
 \bar{\psi}
\left(
\begin{array}{cc}
m_0 & \sigma_k  \partial_k + m_5 \\
- \sigma_k \partial_k + m_5  
-\sigma_k U \partial_k U^{\dag}
& -m_0 
\end{array}
\right)
\psi, 
\nonumber 
\end{align}
where we have renamed $v \to m_0$ and $m\to m_5$. 
Thus, our action can be written as
\begin{align}
\mathcal{L}
&=
\bar{\psi}
\left(
\gamma^k \partial_k
+
m_0 \gamma^0
+
m_5 \gamma_5
+
\gamma^k A_k
+
\gamma^0 \gamma^k B_k
\right)
\psi
\nonumber \\
& \text{where}
\quad
A_k = -B_k = (1/2) U \partial_k U^{\dag}.
\end{align}

The effective action is obtained by 
integrating over fermions,
\begin{align}
\int \mathcal{D}\left[\bar{\psi},\psi\right]
e^{- \int d^3 x\, \mathcal{L}}
= 
e^{-S_{\mathrm{eff}}[U]},
\end{align}
and it can subsequently be 
expanded in powers of $A_{i},B_{i}$ and 
their gradients as
\begin{align}
S_{\mathrm{eff}}
&=
-\mathrm{Tr}\,
\left( G^{-1}_0 -V \right)
\nonumber \\
&=
-\mathrm{Tr}\,
\ln G^{-1}_{0}
+
\sum_{n=1}^{\infty} 
\frac{1}{n}
\mathrm{Tr}\,
\left(G_0 V\right)^n,
\end{align}
where 
$V=
\gamma^i A_i
+
\gamma^0 \gamma^i B_i$,
and
the free propagator is given by
\begin{align}
G_0(k)  &=
-\frac{\gamma^{k}{i}k_{k}+ m_0 \gamma^0+ m_5 \gamma^5 }{k^2 + m^2_0 + m^2_5},
\end{align}
with $M^2=m^2_0+m^2_5$.

To the 3rd order in the gradient expansion, 
the effective action is
[$S^{(n)}_{\mathrm{eff}}
=
(1/n) 
\mathrm{Tr}\,
\left(G_0 V\right)^n$]
\begin{align}
S^{(2)}_{\mathrm{eff}}
&=
\int d^3 x\,
\biggl[
\frac{ \Lambda_{\mbox{\begin{tiny}UV\end{tiny}}}}{3 \pi^2}
\mathrm{tr}\, \left(
A_j A_j
+
B_j B_j
\right)
\nonumber \\
&\quad
-
\frac{{i} m_5}{2\pi |M|}
\epsilon^{ijk}
\mathrm{tr}\, \left(
A_i \partial_j B_k
\right)
+
\frac{m^2_0}{8 \pi |M|}
\mathrm{tr}\, \left(
\partial_j U^{\dag} \partial_j U
\right)
\biggr],
\nonumber \\
S^{(3)}_{\mathrm{eff}}
&=
\int d^3 x\,
\biggl[
-
\frac{ {i}m^3_5}{6\pi |M|^3}
\epsilon^{ijk}
\mathrm{tr}\, \left(
B_i
B_j
B_k
\right)
\nonumber \\
&\quad 
-
\frac{ {i}m_5 (m^2_0+2m^2_5)}{ 4\pi |M|^3}
\epsilon^{ijk}
\mathrm{tr}\, \left(
A_i
A_j
B_k
\right)
\biggr].
\end{align}
By noting that 
\begin{align}
\int d^3 x\,
\epsilon^{ijk} 
\mathrm{tr}\, A_i \partial_j B_k
=
-2 \int d^3 x\,
\epsilon^{ijk} 
\mathrm{tr}\, B_i B_j B_k,
\end{align}
the combined 2nd and 3rd order terms can be written as
\begin{align}
S^{(2)}_{\mathrm{eff}}
+
S^{(3)}_{\mathrm{eff}}
&=
+
\frac{ \Lambda_{\mbox{\begin{tiny}UV\end{tiny}}}}{3 \pi^2}
\int d^3 x\,
\mathrm{tr}\, \left(
A_j A_j
+
B_j B_j
\right)
\nonumber \\
&\quad
+
\frac{m^2_0}{8 \pi |M|}
\int d^3 x\,
\mathrm{tr}\, \left(
\partial_j U^{\dag} \partial_j U
\right)
\nonumber \\
&\quad
-
\frac{1}{8 |M|^3}
\left[
 9 m_5 m^2_0
+4 m^3_5
\right]
2\pi  {i}
\Gamma[U]. 
\label{eq: 2nd and 3rd}
\end{align}

Observe that 
when $m_0=0$, 
if we introduce new gauge fields $A^{\pm}$ by
\begin{align}
A = \frac{1}{2}\left(
A^+ + A^-
\right),
\quad
B = \frac{1}{2}\left(
A^+ - A^-
\right),
\end{align}
the effective action can be written as 
the double Chern-Simons theory: 
\begin{align}
S_{\mathrm{eff}}
&=
+
\frac{ \Lambda_{\mbox{\begin{tiny}UV\end{tiny}}}}{3 \pi^2}
\int d^3 x\,
\mathrm{tr}\, \left(
A_j A_j
+
B_j B_j
\right)
 \\
&\quad
-
\frac{1}{2}
\frac{m_5}{|m_5|}
\frac{{i}}{4\pi}
\sum_{\sigma=\pm}
\sigma 
\int d^3x\,
\epsilon^{ijk}
\nonumber \\
&\quad 
\times 
\mathrm{tr}\, 
\left(
A^{\sigma}_i \partial_j A^{\sigma}_k
+
\frac{2}{3}
A^{\sigma}_i A^{\sigma}_j A^{\sigma}_k
\right).
\nonumber 
\label{double CS}
\end{align}

The effective action 
(\ref{eq: 2nd and 3rd})
[and 
(\ref{double CS})
when $m_0=0$]
that we have derived 
should be regularized
as 
signaled by the appearance of
the term linear in $\Lambda_{\mathrm{UV}}$
in (\ref{eq: 2nd and 3rd})
and 
(\ref{double CS}). 
We introduce the
regularized action
$S^{(R)}_{\mathrm{eff}}$
by the Pauli-Villars 
regularization with the regulator mass $m_R$ as
\begin{align}
&\quad
S^{(R)}_{\mathrm{eff}}
=
S_{\mathrm{eff}}(m_5,m_0)
-
\lim_{|m_R| \to \infty}
S_{\mathrm{eff}}(m_R ,0).
\end{align}
We choose the sign of the regulator mass as
\begin{align}
\mathrm{sgn}\,(m_R)
=
-\mathrm{sgn}\,(m_5)
\end{align}
in which case 
$m_5>0$
is a topological/non-topological phase.
We then conclude
\begin{align}
&
S^{(R)}_{\mathrm{eff}}
=
\frac{m^2_0}{8 \pi |M|}
\int d^3 x\,
\mathrm{tr}\, \left(
\partial_j U^{\dag} \partial_j U
\right)
\\
&
\quad 
+
\left[
\frac{-1}{8 |M|^3}
\left(
 9 m_5 m^2_0
+4 m^3_5
\right)
+\frac{m^3_R}{2 |m_R|^3}
\right]
2\pi  {i}
\Gamma[U].
\nonumber 
\end{align}



\begin{thebibliography}{99}
\bibitem{comment}
One should be warned that
NL$\sigma$Ms might not be useful in three dimensions,
since they are not perturbatively renormalizable. 
On the other hand, it can be renormalizable
non-perturbatively (i.e., in the Wilsonian sense). 
(They are also renormalizable in the large-$N$ expansion,
although in the context of Anderson localization,
$N$ in NL$\sigma$Ms 
is proportional to the number of replicas $N_r$
and 
should be taken to be zero at end of the calculations.)
\end{thebibliography}
\end{document}